\documentclass[a4paper]{article}

\usepackage{a4wide}
\usepackage[UKenglish]{babel}
\usepackage{authblk}
\usepackage{graphicx,subfigure}
	\graphicspath{{./}{figures/}}
\usepackage[colorlinks=true,linkcolor=blue,citecolor=red]{hyperref}
\usepackage{xcolor}
\usepackage{amssymb,amsthm,empheq,bbold}
\usepackage[inline]{enumitem}

\theoremstyle{remark}\newtheorem{remark}{Remark}[section]
\theoremstyle{definition}\newtheorem{assumption}[remark]{Assumption}

\DeclarePairedDelimiter\ave{\langle}{\rangle}
\newcommand{\bA}{\mathbf{A}}
\newcommand{\R}{\mathbb{R}}
\newcommand{\bU}{\mathbf{U}}
\newcommand{\cV}{\mathcal{V}}

\allowdisplaybreaks

\title{Multiscale control of generic second order traffic models by driver-assist vehicles}
\author[1]{Felisia Angela Chiarello}
\author[2]{Benedetto Piccoli}
\author[1]{Andrea Tosin}
\affil[1]{{\footnotesize Department of Mathematical Sciences ``G. L. Lagrange'', Politecnico di Torino, Torino, Italy}}
\affil[2]{{\footnotesize Department of Mathematical Sciences, Rutgers University, Camden NJ, USA}}
 
\date{}

\begin{document}
\maketitle

\begin{abstract}
We study the derivation of generic high order macroscopic traffic models from a follow-the-leader particle description via a kinetic approach. First, we recover a third order traffic model as the hydrodynamic limit of an Enskog-type kinetic equation. Next, we introduce in the vehicle interactions a binary control modelling the automatic feedback provided by driver-assist vehicles and we upscale such a new particle description by means of another Enskog-based hydrodynamic limit. The resulting macroscopic model is now a Generic Second Order Model (GSOM), which contains in turn a control term inherited from the microscopic interactions. We show that such a control may be chosen so as to optimise global traffic trends, such as the vehicle flux or the road congestion, constrained by the GSOM dynamics. By means of numerical simulations, we investigate the effect of this control hierarchy in some specific case studies, which exemplify the multiscale path from the vehicle-wise implementation of a driver-assist control to its optimal hydrodynamic design.
 
\medskip


\medskip

\end{abstract}

\section{Introduction}
Vehicular traffic models incorporating the presence of driver-assist or autonomous vehicles are gaining a lot of momentum. The reason is at least twofold: on one hand, Advanced Driver-Assistance Systems (ADAS), like all technological innovations, call naturally for a quantitative mathematical approach to their understanding and design. On the other hand, ADAS pose new theoretical problems, which motivate interesting developments of mathematical techniques in quite challenging realms such as the one of Artificial Intelligence.

In the literature, several mathematical models at various scales may already be found. Without pretending to be exhaustive, we mention that in~\cite{ntousakis2015TRP} microscopic vehicle-wise control models are reviewed while in~\cite{delis2015CMA} the contribution of adaptive cruise control systems is included in a second order hydrodynamic traffic model. The model is then extended in~\cite{delis2018TRR} to the case of multilane traffic. In~\cite{garavello2020JDE,liard2020PREPRINT} a hybrid microscopic-macroscopic description, inspired by the one introduced for moving bottlenecks~\cite{dellemonache2014JDE,lattanzio2011SIMA} and crowd dynamics~\cite{cristiani2011MMS}, is used to simulate a few individually controlled autonomous vehicles within a continuous traffic stream modelled by the Lighthill-Whitham-Richards traffic equation~\cite{lighthill1955PRSL,richards1956OR}. In~\cite{piccoli2019ZAMP_preprint,tosin2019MMS,tosin2020MCRF} a Boltzmann-type kinetic approach is proposed to account statistically for the presence of driver-assist vehicles in hydrodynamic traffic models and study their impact on mesoscopic traffic features, such as e.g., the local mean speed and speed variability.

Although mathematically different, these models share and convey the idea that driver-assist and autonomous vehicles do not only enhance driver comfort and safety, which were the primary goals for which they were conceived. They also impact in a non-negligible manner on the global traffic flow, to such an extent that one may realistically imagine to take advantage of them as inner \textit{traffic controllers}, as also confirmed by recent field experiments~\cite{stern2018TRC}. In a traffic stream composed mostly of human-manned vehicles but including a certain percentage (the so-called \textit{penetration rate}) of driver-assist vehicles, they make possible an effective \textit{bottom-up} control of traffic trends by exploiting simply the physiological vehicle-to-vehicle interactions. No particular top-down rules imposed by outer traffic controllers are required, whose efficiency would strongly depend also on the hardly controllable voluntary observance by individual drivers.

Inspired by these arguments, in this paper we pursue the research line set up in the already cited papers~\cite{piccoli2019ZAMP_preprint,tosin2019MMS,tosin2020MCRF}. In particular, we aim to derive high order macroscopic traffic models accounting for the presence of driver-assist vehicles to be used as traffic optimisers. The novelties of our contribution with respect to the aforementioned literature may be summarised as follows:
\begin{enumerate*}[label=\roman*)]
\item unlike~\cite{delis2015CMA,delis2018TRR}, we do not postulate the modifications needed in classic hydrodynamic equations of traffic to reproduce the impact of driver-assist vehicles. Instead, we derive them rigorously from an organic upscaling of microscopically controlled particle dynamics;
\item unlike~\cite{garavello2020JDE,liard2020PREPRINT}, we do not regard driver-assist vehicles as point particles, viz. singularities, in a continuous traffic stream. Instead, we derive genuinely macroscopic particle-free models, in which the contribution of driver-assist vehicles is naturally consistent with the upscaling of the whole system;
\item unlike~\cite{piccoli2019ZAMP_preprint,tosin2019MMS,tosin2020MCRF}, we derive hydrodynamic models of order greater than one, which may better account for traffic perturbations and instabilities~\cite{ramadan2020SEMAI-SIMAI}, and we design driver-assist control algorithms having in mind \textit{multiscale} optimisation criteria.
\end{enumerate*}

The mathematical literature offers several techniques to derive macroscopic descriptions of microscopic particle systems, many of them dealing just with vehicular traffic. As an example, we mention micro-macro many particle limits~\cite{colombo2014RSMUP,cristiani2016NHM,difrancesco2017MBE,difrancesco2015ARMA,goatin2017CMS} and mean-field limits~\cite{carrillo2014CISM,carrillo2010MSSET}. Nevertheless, when it comes to controlled particle systems classical approaches become more delicate and difficult due to smoothness issues in the upscaling of the control, see e.g.,~\cite{albi2017AMO,fornasier2014PTRSA}. In this paper, we adopt a ``collisional'' kinetic technique, which is particularly suited to vehicular traffic and is basically free from the technical difficulties just mentioned. Consistently with the classical kinetic approach, we describe the interactions among the vehicles by means of \textit{binary} algebraic rules relating instantaneously the post-interaction states of any \textit{two} interacting vehicles to their pre-interaction states. We notice that binary interactions fit well the \textit{follow-the-leader} particle description classically used in vehicular traffic~\cite{gazis1961OR}. With a probability depending on the penetration rate of driver-assist vehicles, these interactions include furthermore a control term. Therefore, at the particle level we deal with \textit{binary control} problems, which can be easily solved in feedback form: the optimal control can be computed explicitly as a function of the pre-interaction states of the interacting vehicles out of the optimisation of a \textit{binary} cost functional related to their reciprocal distance (\textit{headway}). As a result, we obtain an explicit characterisation of controlled interactions, that we subsequently upscale taking advantage of the classical statistical approach of kinetic theory. In doing so, we adopt in particular an \textit{Enskog-type} kinetic description rather than a more common Boltzmann-type one like in~\cite{piccoli2019ZAMP_preprint,tosin2019MMS,tosin2020MCRF}. Indeed, the Enskog description allows us to properly take into account the fact that the interacting vehicles do not occupy the same spatial position, which is at the basis of the correct reproduction of the anisotropic propagation of traffic waves in high order macroscopic models ~\cite{aw2000SIAP,daganzo1995TR,klar1997JSP}.

The obtained macroscopic description with driver-assist vehicles consists in a second order model belonging to the GSOM class~\cite{appert-rolland2009BOOK,lebacque2007PROCEEDINGS}, which includes as particular cases also the celebrated Aw-Rascle-Zhang model~\cite{aw2000SIAP,zhang2002TRB} and its generalised version (GARZ)~\cite{fan2014NHM}. This model keeps track of the vehicle-wise control in several aspects but notably in a structural parameter of the control, corresponding to a \textit{recommended headway}, which enters the hydrodynamic equations. In order to fix the recommended headway, we propose to set up a further control problem, directly at the macroscopic scale, where this parameter itself plays the role of a control variable for the optimisation of certain cost functionals related to macroscopic traffic features, such as e.g., the vehicle flux and the road congestion. The background idea is to investigate the possibility to design \textit{multiscale} control algorithms for single vehicles which, once embedded in the collective flow, produce bottom-up optimisations of the whole traffic stream.

In more details, the paper is organised as follows. In Section~\ref{sect:h.o._macro_derivation}, we illustrate the general procedure to derive high order macroscopic traffic models from a generic follow-the-leader particle description via the Enskog-type kinetic approach and its hydrodynamic limit. In Section~\ref{sect:GSOM}, we introduce controlled microscopic vehicle interactions and we apply the previous procedure to obtain the corresponding bottom-up controlled macroscopic description in terms of GSOMs. In Section~\ref{sect:hydro_opt}, we tackle the problem of designing the parameters of the vehicle-wise control in such a way to pursue hydrodynamic optimisations. In Section~\ref{sect:numerics}, we show the numerical results produced in some case studies by the macroscopic model with optimally controlled driver-assist vehicles and we compare them with those produced by the more standard GARZ model. As previously anticipated, the latter is in turn a GSOM but in our context we may interestingly interpret it as a model without driver-assist vehicles or alternatively with driver-assist vehicles which do not obey any specific hydrodynamic optimisation criterion. Finally, in Section~\ref{sect:conclusions}, we draw some conclusions and we briefly sketch future research prospects.

\section{Kinetic derivation of generic high order hydrodynamic models}
\label{sect:h.o._macro_derivation}
We begin by showing how hydrodynamic traffic models of order higher than $1$ can be derived from an elementary description of pairwise interactions among the vehicles using a kinetic formalism. This derivation will be the basis to include subsequently a microscopic binary control in the interactions and upscale it at the level of the global macroscopic flow of vehicles.

\subsection{Microscopic Follow-the-Leader description}
\label{sect:FTL}
We begin by considering a generic Follow-the-Leader (FTL) formulation of microscopic traffic dynamics:
\begin{equation}
	\begin{cases}
 		\dot{x}_i=V\!\left(\dfrac{1}{x_{i+1}-x_i},\,\omega_i\right), \\
 		\dot{\omega}_i=0,
	\end{cases}
	\label{eq:FTL}
\end{equation}
where:
\begin{enumerate*}[label=(\roman*)]
\item $x_i,\,\,x_{i+1}\in\R$, $x_i<x_{i+1}$, are the dimensionless positions of two consecutive vehicles in the traffic stream;
\item $\omega_i\in\Omega\subseteq\R_+$ is the so-called \textit{Lagrangian marker}, i.e. a characteristic of the driving style of the drivers, which remains constant in time for each driver. In most cases, $\omega_i$ is interpreted as the maximum speed of the $i$th driver;
\item $V\in [0,\,1]$ is the dimensionless speed of a vehicle expressed as a function of the distance from the leading vehicle and the Lagrangian marker.
\end{enumerate*}
Denoting $s_i:=x_{i+1}-x_{i}\in\R_+$ the headway between the $i$th and $(i+1)$th vehicles, we can restate the model as
\begin{equation}
	\begin{cases}
 		\dot{s}_i=V\!\left(\dfrac{1}{s_{i+1}},\,\omega_{i+1}\right)-V\!\left(\dfrac{1}{s_i},\,\omega_{i}\right), \\
 		\dot{\omega}_i=0.
	\end{cases}
	\label{eq:FTL.s}
\end{equation}

\begin{assumption}
We assume that:
\begin{enumerate}[label=(\roman*)]
\item \label{item:partials.V} $\partial_sV(\frac{1}{s},\,\omega)>0$, $\forall\,(s,\,\omega)\in\R_+\times\Omega$ (\footnote{We point out that, here and henceforth, the notation $\partial_sV(\frac{1}{s},\,\omega)$ stands for
$$ \partial_sV\!\left(\frac{1}{s},\,\omega\right):=(\partial_sV)\!\left(\frac{1}{s},\,\omega\right)=-\frac{1}{s^2}(\partial_\sigma V)\!\left(\frac{1}{s},\,\omega\right), $$
where $\sigma$ denotes the first variable of the function $V$. In practice, we consider $V=V(\sigma,\,\omega)$ along with the composition $\sigma(s)=\frac{1}{s}$ and we take the derivatives accordingly.});
\item \label{item:V.bound.lin} $\exists\,C>0$ such that $V\!\left(\frac{1}{s},\,\omega\right)\leq Cs$, $\forall\,(s,\,\omega)\in\R_+\times\Omega$.
\end{enumerate}
\label{ass:V}
\end{assumption}

\begin{remark}
In different derivations of high order macroscopic traffic models from the FTL description~\eqref{eq:FTL}, see e.g.,~\cite{chiarello2020SIAM,fan2014NHM}, further assumptions are made on the function $V$, which however are not needed in the present context.

A function $V$ complying with Assumption~\ref{ass:V} is
\begin{equation}
	V\!\left(\frac{1}{s},\,\omega\right)=\frac{\omega s}{a+s}
	\label{eq:V.FTL}
\end{equation}
with $a>0$ constant and $\omega\in\Omega:=[0,\,1]$. This function is motivated by the form of the speed of vehicles in classical FTL models, cf. e.g.,~\cite{piccoli2019ZAMP_preprint,tosin2020SEMA-SIMAI}.
\end{remark}

Following~\cite{carrillo2010SIMA}, we now use~\eqref{eq:FTL.s} to obtain a set of binary interaction rules between any two consecutive vehicles. Specifically, we approximate~\eqref{eq:FTL.s} with the forward Euler formula in a small time interval $\gamma>0$, understood as the reaction time of the drivers. Denoting by $s:=s_i(t)$, $s_\ast:=s_{i+1}(t)$ the pre-interaction headways and by $s':=s_i(t+\gamma)$, $s_\ast':=s_{i+1}(t+\gamma)$ the post-interaction headways, and using an analogous notation for the Lagrangian markers, we get
\begin{equation}
	s'=s+\gamma\left[V\!\left(\frac{1}{s_\ast},\,\omega_\ast\right)-V\!\left(\frac{1}{s},\,\omega\right)\right], \qquad
		\omega'=\omega,
	 \label{eq:interactions}
\end{equation}
that we may further complement with $s_\ast'=s_\ast$ to express the anisotropy of vehicle interactions, in particular the fact that the leading vehicle is not affected by the rear vehicle.

For physical consistency, the interaction~\eqref{eq:interactions} has to guarantee $s'\geq 0$ for all $s,\,s_\ast\geq 0$ and all $\omega,\,\omega_\ast\in\Omega$. Thanks to Assumption~\ref{ass:V}\ref{item:V.bound.lin}, we easily see that this condition is met if $\gamma\leq\frac{1}{C}$.

\subsection{Enskog-type kinetic description}
\label{sect:Enskog_without_control}
The aggregate outcome of the microscopic binary interactions~\eqref{eq:interactions} may be investigated through a kinetic approach upon introducing the distribution function $f=f(t,\,x,\,s,\,\omega)\geq 0$ such that $f(t,\,x,\,s,\,\omega)\,dx\,ds\,d\omega$ gives, at time $t>0,$ the fraction of vehicles located in the interval $[x,\,x+dx]$ with headway comprised in $[s,\,s+ds]$ and Lagrangian marker in $[\omega,\,\omega+d\omega]$.

In this work, we assume that $f$ satisfies an \textit{Enskog-type} kinetic equation rather than a more classical Boltzmann-type equation. The inspiration comes from~\cite{dimarco2020JSP,herty2007NHM,klar1997JSP}, where it is stressed that traffic models derived from a Boltzmann-type kinetic description cannot reproduce backward wave propagation because in a Boltzmann-type equation the interacting vehicles are assumed to occupy the same space position. Conversely, in an Enskog-type kinetic description they are assumed to occupy two different positions, which in our case is also particularly consistent with the fact that their microscopic state includes the headway, viz. the reciprocal distance. We write therefore:
\begin{equation}
	\partial_tf+V\!\left(\frac{1}{s},\,\omega\right)\partial_xf=Q_\mathrm{E}(f,\,f),
	\label{eq:Enskog.strong}
\end{equation}
where $Q_\mathrm{E}(f,\,f)$ is the Enskog collision operator. The precise definition of $Q_\mathrm{E}(f,\,f)$ is better given in weak form, i.e. through its action on an arbitrary macroscopic observable (test function) $\phi=\phi(s,\,\omega)$:
\begin{equation}
	(Q_E(f,\,f),\,\phi):=\frac{1}{2}\int_{\Omega^2}\int_{\R_+^2}(\phi(s',\,\omega')-\phi(s,\,\omega))f(t,\,x,\,s,\,\omega)f(t,\,x+s,\,s_\ast,\,\omega_\ast)\,ds\,ds_\ast\,d\omega\,d\omega_\ast,
	\label{eq:QE.weak}
\end{equation}
where $s',\,\omega'$ are given by~\eqref{eq:interactions}. Notice that the two distribution functions describing the interacting vehicles are computed in $x$ and $x+s$, respectively. Indeed, if $s$ is the headway of the rear vehicle located in $x$ then the leading vehicle is located in $x+s$.

In order to make~\eqref{eq:Enskog.strong},~\eqref{eq:QE.weak} more amenable to analytical investigations, it is useful to approximate
\begin{equation}
	f(t,\,x+s,\,s_\ast,\,\omega_\ast)\approx f(t,\,x,\,s_\ast,\,\omega_\ast)+\partial_xf(t,\,x,\,s_\ast,\,\omega_\ast)s,
	\label{eq:f.Taylor}
\end{equation}
which, for $s$ sufficiently small, coincides with the first order Taylor expansion of $f$ in $x$. Then~\eqref{eq:QE.weak} takes the form
\begin{align}
	\begin{aligned}[b]
		(Q_\mathrm{E}(f,\,f),\,\phi) &= \frac{1}{2}\int_{\Omega^2}\int_{\R_+^2}(\phi(s',\,\omega')-\phi(s,\,\omega))f(t,\,x,\,s,\,\omega)f(t,\,x,\,s_\ast,\,\omega_\ast)\,ds\,ds_\ast\,d\omega\,d\omega_\ast \\
		&\phantom{=} +\frac{1}{2}\int_{\Omega^2}\int_{\R_+^2}(\phi(s',\,\omega')-\phi(s,\,\omega))f(t,\,x,\,s,\,\omega)\partial_xf(t,\,x,\,s_\ast,\,\omega_\ast)s\,ds\,ds_\ast\,d\omega\,d\omega_\ast \\
		&=: (Q(f,\,f),\,\phi)+(Q(f,\,s\partial_xf),\,\phi).
	\end{aligned}
	\label{eq:QE.Taylor}
\end{align}
The first term on the right-hand side, i.e. $Q(f,\,f)$, is now a classical Boltzmann-type collision operator with the two distribution functions computed in the same point $x$. The second term $Q(f,\,s\partial_xf)$ is instead a first order correction, which will be fundamental to recover consistent macroscopic models.

The passage from~\eqref{eq:Enskog.strong} to a macroscopic traffic description is performed via the so-called \textit{hydrodynamic limit}. Let $0<\eta\ll 1$ be a small scale parameter (the analogous of the Knudsen number in gas and fluid dynamics) and let us introduce the following hyperbolic scaling of time and space:
\begin{equation}
	t\to\frac{t}{\eta}, \qquad x\to\frac{x}{\eta},
	\label{eq:hyp_scaling}
\end{equation}
which formalises the passage from microscopic to macroscopic time and space scales. Then~\eqref{eq:Enskog.strong},~\eqref{eq:QE.Taylor} become
\begin{equation}
	\partial_tf+V\!\left(\frac{1}{s},\,\omega\right)\partial_xf=\frac{1}{\eta}Q_\mathrm{E}(f,\,f)
	\label{eq:Enskog.strong.scaled}
\end{equation}
and
\begin{align*}
	(Q_\mathrm{E}(f,\,f),\,\phi) &= \frac{1}{2}\int_{\Omega^2}\int_{\R_+^2}(\phi(s',\,\omega')-\phi(s,\,\omega))f(t,\,x,\,s,\,\omega)f(t,\,x,\,s_\ast,\,\omega_\ast)\,ds\,ds_\ast\,d\omega\,d\omega_\ast \\
	&\phantom{=} +\frac{\eta}{2}\int_{\Omega^2}\int_{\R_+^2}(\phi(s',\,\omega')-\phi(s,\,\omega))f(t,\,x,\,s,\,\omega)\partial_xf(t,\,x,\,s_\ast,\,\omega_\ast)s\,ds\,ds_\ast\,d\omega\,d\omega_\ast \\
	&= (Q(f,\,f),\,\phi)+\eta(Q(f,\,s\partial_xf),\,\phi)
\end{align*}
(for simplicity, we still denote by $f=f(t,\,x,\,s,\,\omega)$ the distribution function in the scaled time and space variables). Hence, $Q_\mathrm{E}(f,\,f)=Q(f,\,f)+\eta Q(f,\,s\partial_xf)$, which plugged into~\eqref{eq:Enskog.strong.scaled} yields
\begin{equation}
	\partial_tf+V\!\left(\frac{1}{s},\,\omega\right)\partial_xf=\frac{1}{\eta}Q(f,\,f)+Q(f,\,s\partial_xf).
	\label{eq:Enskog_scaled}
\end{equation}
Owing to the smallness of $\eta$, this equation can be split in two contributions. On one hand, \textit{local} interactions among the vehicles, which take place on a microscopic (quick) time scale and reach rapidly the equilibrium:
\begin{equation}
	\partial_tf=Q(f,\,f)
	\label{eq:interaction_step}
\end{equation}
(we have scaled the time back to the microscopic scale as $t\to\eta t$ using the factor $\frac{1}{\eta}$ in front of the collision operator); on the other hand, a transport of the local equilibrium distribution generated by~\eqref{eq:interaction_step} on a \textit{hydrodynamic} (slow) time scale:
\begin{equation}
	\partial_tf+V\!\left(\frac{1}{s},\,\omega\right)\partial_xf=Q(f,\,s\partial_xf).
	\label{eq:transport_step}
\end{equation}
Here, we use the \textit{local Maxwellian}, viz. the equilibrium distribution produced by~\eqref{eq:interaction_step}, to obtain the macroscopic evolution of the hydrodynamic parameters locally conserved by the interactions.

\subsection{Hydrodynamic limit}
\subsubsection{Local Maxwellian}
\label{sect:local_Maxwellian.uncontrolled}
The first step of the strategy just outlined is the study of the local equilibrium distribution resulting from~\eqref{eq:interaction_step}. In weak form,~\eqref{eq:interaction_step} reads
\begin{align}
	\begin{aligned}[b]
		\partial_t & \int_\Omega\int_{\R_+}\phi(s,\,\omega)f(t,\,x,\,s,\,\omega)\,ds\,d\omega \\
		&= \frac{1}{2}\int_{\Omega^2}\int_{\R_+^2}(\phi(s',\,\omega')-\phi(s,\,\omega))f(t,\,x,\,s,\,\omega)f(t,\,x,\,s_\ast,\,\omega_\ast)\,ds\,ds_\ast\,d\omega\,d\omega_\ast.
	\end{aligned}
	\label{eq:interaction_step.weak}
\end{align}
Choosing $\phi(s,\,\omega)=1$ and defining the macroscopic density of the vehicles in $x$ as
$$ \rho(t,\,x):=\int_\Omega\int_{\R_+}f(t,\,x,\,s,\,\omega)\,ds\,d\omega $$
we immediately observe that $\rho$ is conserved in time by the local interactions. Likewise, choosing $\phi(s,\,\omega)=s$ and defining the mean headway in $x$ as
$$ h(t,\,x):=\frac{1}{\rho(t,\,x)}\int_\Omega\int_{\R_+}sf(t,\,x,\,s,\,\omega)\,ds\,d\omega $$
we see that also $h$ is locally conserved in time owing to~\eqref{eq:interactions}. Finally, choosing $\phi(s,\,\omega)=\omega$ and defining the mean Lagrangian marker in $x$ as
$$ w(t,\,x):=\frac{1}{\rho(t,\,x)}\int_\Omega\int_{\R_+}\omega f(t,\,x,\,s,\,\omega)\,ds\,d\omega $$
we obtain from~\eqref{eq:interactions} that also $w$ is locally conserved in time. We conclude that $\phi(s,\,\omega)=1,\,s,\,\omega$ are ``collisional invariants'' and therefore that the local Maxwellian will be parametrised by the hydrodynamic quantities $\rho$, $h$, $w$.

More in general, choosing in~\eqref{eq:interaction_step.weak} a macroscopic observable $\phi(s,\,\omega)=\psi(\omega)$ independent of $s$ and using~\eqref{eq:interactions} we deduce
$$ \partial_t\int_\Omega\int_{\R_+}\psi(\omega)f(t,\,x,\,s,\,\omega)\,ds\,d\omega=0, $$
i.e. the whole marginal distribution of $\omega$ is locally constant in time. Consequently, the local Maxwellian should be parametrised by all the statistical moments of the $\omega$-marginal. To avoid an infinite proliferation of hydrodynamic parameters, we assume that the $\omega$-marginal is of the form $\delta(\omega-w)$, where $\delta$ denotes the Dirac delta, so that all its moments can be expressed in terms of $w$. This leads us to consider a distribution function $f$ of the form
\begin{equation}
	f(t,\,x,\,s,\,\omega):=\rho(x)g_{h(x)}(t,\,s)\delta(\omega-w(x)),
	\label{eq:f.interaction_step}
\end{equation}
where $g_h$ is the marginal of $s$ parametrised by the conserved mean headway $h$:
$$ \int_{\R_+}g_h(t,\,s)\,ds=1, \qquad \int_{\R_+}sg_h(t,\,s)\,ds=h \qquad \forall\,t\geq 0. $$
We point out that in~\eqref{eq:f.interaction_step} we have omitted the dependence of $\rho$, $h$, $w$ on $t$ because these hydrodynamic parameters are constant on the time scale of the microscopic interactions.

Plugging~\eqref{eq:f.interaction_step} into~\eqref{eq:interaction_step.weak} and choosing a macroscopic observable $\phi(s,\,\omega)=\varphi(s)$ independent of $\omega$ we deduce the following equation for $g_h$:
\begin{align}
	\begin{aligned}[b]
		&\frac{d}{dt}\int_{\R_+}\varphi(s)g_h(t,\,s)\,ds \\
		&= \frac{\rho}{2}\int_{\Omega^2}\int_{\R_+^2}(\varphi(s')-\varphi(s))g_h(t,\,s)g_h(t,\,s_\ast)\delta(\omega-w)\delta(\omega_\ast-w)\,ds\,ds_\ast\,d\omega\,d\omega_\ast,
	\end{aligned}
	\label{eq:gh}
\end{align}
which, in view of~\eqref{eq:interactions}, admits
\begin{equation}
	g_h^\infty(s):=\delta(s-h)
	\label{eq:g_h^infty}
\end{equation}
as an equilibrium distribution. Indeed, a direct calculation shows that such a $g_h^\infty$ makes the right-hand side of~\eqref{eq:gh} vanish. In general,~\eqref{eq:g_h^infty} may not be the only possible equilibrium distribution of~\eqref{eq:gh} under the interaction rules~\eqref{eq:interactions} due to the arbitrariness of the speed function $V$. In the following we prove however that~\eqref{eq:g_h^infty} is the unique equilibrium distribution at least in a particular regime of the parameters of the interactions~\eqref{eq:interactions}, which allows us to identify a ``universal'' trend substantially independent of $V$.

Let us consider \textit{quasi-invariant interactions}, namely interactions which induce a small change of the microscopic state of the vehicles. This concept is inspired by the \textit{grazing collisions} of the classical kinetic theory~\cite{villani1998PhD,villani1998ARMA} and has been introduced in the kinetic theory of multi-agent system in~\cite{cordier2005JSP}. In~\eqref{eq:interactions}, this is the case if e.g., $V(\frac{1}{s_\ast},\,\omega_\ast)-V(\frac{1}{s},\,\omega)$ is small so that $s'\approx s$. Let us assume that $V$ is parametrised by a parameter $\epsilon>0$ such that
$$ V\!\left(\frac{1}{s},\,\omega\right)\sim\epsilon c(\omega)s \qquad \text{for } \epsilon\to 0^+, $$
where $c(\omega)\geq 0$ denotes a function of $\omega$. This implies that there exists a function $\cV_\epsilon(s)$ such that $\cV_\epsilon(s)\to 1$ when $\epsilon\to 0^+$ and
\begin{equation}
	V\!\left(\frac{1}{s},\,\omega\right)=\epsilon c(\omega)s\cV_\epsilon(s).
	\label{eq:V.small_epsilon}
\end{equation}
We will further assume that $\cV_\epsilon(s)$ is bounded for all $\epsilon>0$ and $s\in\R_+$. For example, if we let $a=\frac{1}{\epsilon}$ then the function $V$ given in~\eqref{eq:V.FTL} satisfies~\eqref{eq:V.small_epsilon} with $c(\omega)=\omega$ and $\cV_\epsilon(s)=\frac{1}{1+\epsilon s}$.

Obviously, with the sole assumption of small $\epsilon$ we cannot observe any interesting universal trend of the interactions towards the equilibrium. Indeed, in the limit $\epsilon\to 0^+$ we simply get $s'=s$ in~\eqref{eq:interactions}, which implies definitively a constant solution $f$ to~\eqref{eq:interaction_step.weak} coinciding with the arbitrarily chosen initial local distribution. To compensate for the smallness of $\epsilon$ we increase simultaneously the frequency of the interactions as $\frac{1}{\epsilon}$, so as to balance the small transfer of microscopic state from one vehicle to another in a single interaction with a high number of such interactions per unit time. Hence, in the quasi-invariant regime we consider~\eqref{eq:gh} in the form
\begin{align}
	\begin{aligned}[b]
		&\frac{d}{dt}\int_{\R_+}\varphi(s)g_h(t,\,s)\,ds \\
		&= \frac{\rho}{2\epsilon}\int_{\Omega^2}\int_{\R_+^2}(\varphi(s')-\varphi(s))g_h(t,\,s)g_h(t,\,s_\ast)\delta(\omega-w)\delta(\omega_\ast-w)\,ds\,ds_\ast\,d\omega\,d\omega_\ast.
	\end{aligned}
	\label{eq:gh.quasi_inv}
\end{align}
Notice that the scaling of the interaction frequency does not affect either the equilibrium distributions or the conservation of $h$. The first statistical moment of $g_{h(x)}$ which in general is not conserved by the microscopic interactions is still the second moment, namely the \textit{energy}
$$ E(t):=\int_{\R_+}s^2g_h(t,\,s)\,ds, $$
whose trend is provided by~\eqref{eq:gh.quasi_inv} with $\varphi(s)=s^2$:
\begin{align*}
	\frac{dE}{dt} &= \frac{\gamma\rho}{\epsilon}\int_{\R_+^2}s\left[V\!\left(\frac{1}{s_\ast},\,w\right)-V\!\left(\frac{1}{s},\,w\right)\right]g_h(t,\,s)g_h(t,\,s_\ast)\,ds\,ds_\ast \\
	&\phantom{=} +\frac{\gamma\rho}{2\epsilon}\int_{\R_+^2}{\left[V\!\left(\frac{1}{s_\ast},\,w\right)-V\!\left(\frac{1}{s},\,w\right)\right]}^2g_h(t,\,s)g_h(t,\,s_\ast)\,ds\,ds_\ast.
\end{align*}
Recalling~\eqref{eq:V.small_epsilon}, this yields
\begin{align*}
	\frac{dE}{dt} &= \gamma\rho c(w)\int_{\R_+^2}s(s_\ast\cV_\epsilon(s_\ast)-s\cV_\epsilon(s))g_h(t,\,s)g_h(t,\,s_\ast)\,ds\,ds_\ast \\
	&\phantom{=} +\frac{\gamma\epsilon}{2}\rho c^2(w)\int_{\R_+^2}{(s_\ast\cV_\epsilon(s_\ast)-s\cV_\epsilon(s))}^2g_h(t,\,s)g_h(t,\,s_\ast)\,ds\,ds_\ast
\end{align*}
and finally, passing to the limit $\epsilon\to 0^+$ by dominated convergence to obtain a universal trend for small $\epsilon$,
\begin{equation}
	\frac{dE}{dt}=\gamma\rho c(w)(h^2-E).
	\label{eq:E}
\end{equation}
From this equation we deduce $E\to h^2$ for $t\to +\infty$, thus the variance $E-h^2$ of the equilibrium distribution $g_h^\infty$ vanishes asymptotically. This proves that~\eqref{eq:g_h^infty} is the unique distribution towards which the system converges for large times in the quasi-invariant regime.

Motivated by these arguments, we finally consider the following local Maxwellian as the result of the local interaction step~\eqref{eq:interaction_step}:
\begin{equation}
	M_{\rho,h,w}(s,\,\omega)=\rho\delta(s-h)\otimes\delta(\omega-w).
	\label{eq:M}
\end{equation}

\subsubsection{Macroscopic equations}
\label{sect:macro_eq.uncontrolled}
Macroscopic equations are obtained by plugging the local Maxwellian~\eqref{eq:M} into~\eqref{eq:transport_step} to determine evolution equations for the hydrodynamic parameters $\rho$, $h$, $w$:
\begin{equation}
	\partial_tM_{\rho,h,w}+V\!\left(\frac{1}{s},\,\omega\right)\partial_xM_{\rho,h,w}=Q(M_{\rho,h,w},\,s\partial_xM_{\rho,h,w}).
	\label{eq:transport_step.M}
\end{equation}
We stress that here we need to restore the dependence of the hydrodynamic parameters on time because they are in general not constant on the time scale of the hydrodynamic transport.

Writing~\eqref{eq:transport_step.M} in weak form and using~\eqref{eq:M} we get
$$ \partial_t\left(\rho\phi(h,\,w)\right)+\partial_x\!\left(\rho\phi(h,\,w)V\!\left(\frac{1}{h},\,w\right)\right)
	=\frac{\gamma}{2}\rho^2h\partial_xV\!\left(\frac{1}{h},\,w\right)\partial_s\phi(h,\,w), $$
whence for $\phi(s,\,\omega)=1,\,\omega,\,s$ (the collisional invariants) we obtain the \textit{third order} hydrodynamic system
\begin{equation}
	\begin{cases}
 		\partial_t\rho+\partial_x\!\left(\rho V\!\left(\dfrac{1}{h},\,w\right)\right)=0 \\[4mm]
 		\partial_t(\rho w)+\partial_x\!\left(\rho wV\!\left(\dfrac{1}{h},\,w\right)\right)=0 \\[4mm]
 		\partial_t(\rho h)+\partial_x\!\left(\rho hV\!\left(\dfrac{1}{h},\,w\right)\right)=\dfrac{\gamma}{2}\rho^2h\partial_xV\!\left(\dfrac{1}{h},\,w\right).
	\end{cases}
	\label{eq:thirdorderwithoutcontrol}
\end{equation}
The first two equations express a classical conservative transport of the density of the vehicles and of their mean Lagrangian marker by the velocity field $V$. The third equation deserves instead a couple of further comments. First, this additional equation is present because the microscopic interactions~\eqref{eq:interactions} conserve locally also $h$. Second, it expresses a balance and not a conservation, i.e. the right-hand side is not zero, because of the non-local correction to the vehicle interactions included in the Enskog collision operator~\eqref{eq:QE.Taylor}. Third order models were already occasionally proposed in the traffic literature, see~\cite{helbing1995PRE} for an example, however not within an organic derivation from microscopic principles like in this case.

System~\eqref{eq:thirdorderwithoutcontrol} can be written in quasilinear vector form as
$$ \partial_t\bU+\bA(\bU)\partial_x\bU=\mathbf{0}, $$
with $\bU:=(\rho,\,w,\,h)^T$ and
$$	\bA(\bU):=
	\begin{pmatrix}
		V(\frac{1}{h},\,w) & \rho\partial_\omega V(\frac{1}{h},\,w) & \rho\partial_sV(\frac{1}{h},\,w) \\
 		0 & V(\frac{1}{h},\,w) & 0 \\
 		0 & -\frac{\gamma}{2}\rho h\partial_\omega V(\frac{1}{h},\,w) & V(\frac{1}{h},\,w)-\frac{\gamma}{2}\rho h\partial_sV(\frac{1}{h},\,w)
 	\end{pmatrix},
$$
cf. Assumption~\ref{ass:V} for the correct interpretation of $\partial_sV$. The eigenvalues $\lambda_1,\,\lambda_2,\,\lambda_3$ and eigenvectors $r_1,\,r_2,\,r_3$ of this matrix are 
$$ \lambda_1=\lambda_2=V\!\left(\frac{1}{h},\,w\right) \quad \text{with} \quad r_1=(1,\,0,\,0), \quad
	r_2=\left(0,\,\partial_sV\!\left(\frac{1}{h},\,w\right),\,\partial_\omega V\!\left(\frac{1}{h},\,w\right)\right) $$
and
$$ \lambda_3=V\!\left(\frac{1}{h},\,w\right)-\frac{\gamma}{2}\rho h\partial_sV\!\left(\frac{1}{h},\,w\right)
	\quad \text{with} \quad r_3=\left(1,\,0,\,-\frac{\gamma}{2}h\right). $$
Since the eigenvalues are real and $\bA(\bU)$ is diagonalisable, system~\eqref{eq:thirdorderwithoutcontrol} is hyperbolic. Nevertheless, since $\lambda_1=\lambda_2$ it is not strictly hyperbolic. Furthermore, under Assumption~\ref{ass:V}\ref{item:partials.V} it results $\lambda_3<\lambda_1=\lambda_2=V$, therefore no characteristic speed is greater than the flow speed. Hence~\eqref{eq:thirdorderwithoutcontrol} complies with the Aw-Rascle consistency condition~\cite{aw2000SIAP}.
The first and second characteristic fields are linearly degenerate: $\nabla\lambda_1\cdot r_1=\nabla\lambda_2\cdot r_2=0$, thus the associated waves are contact discontinuities. Conversely, the third characteristic field is genuinely nonlinear: $\nabla\lambda_3\cdot r_3\neq 0$, hence the associated waves are either shocks or rarefactions.

\section{Derivation of GSOM with driver-assist vehicles}
\label{sect:GSOM}
In this section, we take advantage of the procedure illustrated in Section~\ref{sect:h.o._macro_derivation} to derive similar macroscopic traffic models incorporating the presence of driver-assist vehicles. At the microscopic scale, the latter are regarded as special vehicles equipped with automatic feedback controllers, which respond locally to the actions of the human drivers with the aim of optimising a certain cost functional in each binary interaction. We anticipate that the introduction of controlled vehicles will give rise to second (rather than third) order hydrodynamic models.

\subsection{Microscopic binary control}
\label{sect:micro_control}
To implement the presence of driver-assist vehicles, we restate the interaction rules~\eqref{eq:interactions} as follows:
\begin{equation}
	s'=s+\gamma\left[V\!\left(\frac{1}{s_\ast},\,\omega_\ast\right)-V\!\left(\frac{1}{s},\,\omega\right)+\Theta u\right], \qquad
		\omega'=\omega.
	\label{eq:interactionscontrol} 
\end{equation}
Here, $u\in\R$ denotes the control applied to the dynamics of a driver-assist vehicle and $\Theta\in\{0,\,1\}$ is a Bernoulli random variable expressing the fact that a randomly chosen vehicle may or may not be equipped with a driver-assist technology with a certain probability. In particular, by letting
$$ \operatorname{Prob}(\Theta=1)=p, \qquad \operatorname{Prob}(\Theta=0)=1-p $$
we mean that $p\in [0,\,1]$ is the percentage of driver-assist vehicles in the traffic stream, namely the so-called \textit{penetration rate}.

Aiming at \textit{collision avoidance}, the control $u$ is chosen so as to minimise the following cost functional:
\begin{equation}
	J(s',\,u):=\frac{1}{2}\Bigl({\left(s_d(\rho,\,w)-s'\right)}^2+\nu u^2\Bigr),
	\label{eq:J}
\end{equation}
where $s_d(\rho,\,w)\geq 0$ is a recommended headway that vehicles should maintain depending on the local hydrodynamic parameters $\rho,\,w$ and $\nu>0$ is a penalisation parameter (cost of the control). By minimising the functional~\eqref{eq:J}, the control $u$ tries to align the headway of the vehicle to the recommended one, thereby implementing a form of collision avoidance. The optimal control $u^\ast$ is chosen as
$$ u^\ast:=\operatornamewithlimits{arg\,min}_{u\in\mathcal{U}} J(s',\,u) $$
subject to~\eqref{eq:interactionscontrol}, where $\mathcal{U}=\{u\in\R\,:\,s'\geq 0\}$ is the set of the admissible controls.

Plugging the constraint~\eqref{eq:interactionscontrol} into~\eqref{eq:J} and equating to zero the derivative with respect to $u$, we deduce the following optimality condition:
$$ \gamma\Theta\left\{s-s_d(\rho,\,w)+\gamma\left[V\!\left(\frac{1}{s_\ast},\,\omega_\ast\right)-V\!\left(\frac{1}{s},\,\omega\right)\right]\right\}+\left(\nu+\gamma^2\Theta^2\right)u^\ast=0 $$
yielding
\begin{equation}
	u^\ast=\frac{\Theta\gamma}{\nu+\Theta^2\gamma^2}(s_d(\rho,\,w)-s)
		-\frac{\Theta\gamma^2}{\nu+\Theta^2\gamma^2}\left[V\!\left(\frac{1}{s_\ast},\,\omega_\ast\right)-V\!\left(\frac{1}{s},\,\omega\right)\right].
	\label{eq:u^ast}
\end{equation}
Notice that $u^\ast$ is a \textit{feedback control} because it is a function of the pre-interaction states $s$, $s_\ast$, $\omega$, $\omega_\ast$ of the vehicles. This allows us to plug it straightforwardly into~\eqref{eq:interactionscontrol}, whence we obtain the following controlled binary interactions:
\begin{equation} 
	s'=s+\frac{\gamma}{\nu+\Theta^2\gamma^2}\left\{\nu\left[V\!\left(\frac{1}{s_\ast},\,\omega_\ast\right)-V\!\left(\frac{1}{s},\,\omega\right)\right]
		+\Theta^2\gamma\left(s_d(\rho,\,w)-s\right)\right\}, \qquad \omega'=\omega.
	\label{eq:controlled_interactions} 
\end{equation}

Finally, we check that $u^\ast\in\mathcal{U}$, which amounts to checking the physical admissibility of the controlled interaction~\eqref{eq:controlled_interactions}. Recalling Assumption~\ref{ass:V}\ref{item:V.bound.lin} and considering that $0\leq\Theta^2\leq 1$, we easily see that the condition $s'\geq 0$ is fulfilled if e.g.,
$$ \nu\geq\frac{\gamma^2}{1-C\gamma} $$
under the further restriction $\gamma\leq\frac{1}{C}$ already established in Section~\ref{sect:FTL}. This condition implies that there is a physiological lower bound on the cost of the implementation of the driver-assist control, which cannot be assumed too cheap.

\begin{remark}
If $\nu\to +\infty$ then $u^\ast=0$. In this case, from~\eqref{eq:controlled_interactions} we recover the uncontrolled interaction rules~\eqref{eq:interactions}. Another case in which we obtain~\eqref{eq:interactions} from~\eqref{eq:controlled_interactions} is if $\Theta=0$, which corresponds to a vehicle without driver-assist control.
\end{remark}

\subsection{Enskog-type kinetic description and hydrodynamic limit}
\label{sect:Enskog_with_control}
The Enskog-type description is the same as the one discussed in Section~\ref{sect:Enskog_without_control} but for the fact that the collision operator $Q_E(f,\,f)$ takes now into account also the presence of the random parameter $\Theta$ in the interaction rules~\eqref{eq:controlled_interactions}. Specifically, the generalisation of~\eqref{eq:QE.weak} to the present case reads
$$ (Q_E(f,\,f),\,\phi)=\frac{1}{2}\ave*{\int_{\Omega^2}\int_{\R_+^2}(\phi(s',\,\omega')-\phi(s,\,\omega))f(t,\,x,\,s,\,\omega)f(t,\,x+s,\,s_\ast,\,\omega_\ast)\,ds\,ds_\ast\,d\omega\,d\omega_\ast}, $$
where $\ave{\cdot}$ denotes the expectation with respect to the law of $\Theta$.

The same expansion~\eqref{eq:f.Taylor} followed by the hyperbolic scaling~\eqref{eq:hyp_scaling} leads again to~\eqref{eq:Enskog_scaled}, where the Boltzmann-type collision operator $Q(f,\,f)$ includes in turn the expectation with respect to $\Theta$:
$$ (Q(f,\,f),\,\phi)=\frac{1}{2}\ave*{\int_{\Omega^2}\int_{\R_+^2}(\phi(s',\,\omega')-\phi(s,\,\omega))f(t,\,x,\,s,\,\omega)f(t,\,x,\,s_\ast,\,\omega_\ast)\,ds\,ds_\ast\,d\omega\,d\omega_\ast}. $$
Choosing $\phi(s,\,\omega)=1$ and $\phi(s,\,\omega)=\psi(\omega)$ (a function of $\omega$ alone) and using~\eqref{eq:controlled_interactions} we see that
$$ (Q(f,\,f),\,1)=(Q(f,\,f),\,\psi(\omega))=0, $$
hence the mass of the vehicles as well as any statistical moment of the $\omega$-marginal are locally conserved by the controlled interactions. Conversely, choosing $\phi(s,\,\omega)=s$ we discover
$$ (Q(f,\,f),\,s)=\frac{p\gamma^2\rho^2}{2\left(\nu+\gamma^2\right)}\left(s_d(\rho,\,w)-h\right), $$
meaning that on the scale of the local interactions the evolution of the mean headway is ruled by
\begin{equation}
	\frac{dh}{dt}=\frac{p\gamma^2\rho}{2\left(\nu+\gamma^2\right)}\left(s_d(\rho,\,w)-h\right).
	\label{eq:h.sd}
\end{equation}
We point out that in this equation we are omitting the dependence of $\rho,\,h,\,w$ on $x$ for brevity, considering that for local interactions $x$ is a parameter. Moreover, here $\rho,\,w$ have to be regarded as constant with respect to $t$ in view of the conservations discussed above. From~\eqref{eq:h.sd} we deduce that $h$ is no longer conserved by the interactions~\eqref{eq:controlled_interactions} and, in particular, that it converges exponentially fast in time to $s_d(\rho,\,w)$ at a rate proportional to the penetration rate $p$.

Out of these arguments, we conclude that an admissible form of the kinetic distribution function in the local interaction step is
$$ f(t,\,x,\,s,\,\omega)=\rho(x)g(t,\,s)\delta(\omega-w(x)), $$
where the $\omega$-marginal is chosen based on the same considerations as in Section~\ref{sect:local_Maxwellian.uncontrolled}. Conversely, the distribution $g$ now satisfies only the normalisation condition
$$ \int_{\R_+}g(t,\,s)\,ds=1 \qquad \forall\,t\geq 0 $$
because the mean headway is not conserved by the controlled interactions. Similarly to~\eqref{eq:gh}, the evolution equation for $g$ can then be written in the form
\begin{align*}
	&\frac{d}{dt}\int_{\R_+}\varphi(s)g(t,\,s)\,ds \\
	&=\frac{\rho}{2}\ave*{\int_{\Omega^2}\int_{\R_+^2}\left(\varphi(s')-\varphi(s)\right)g(t,\,s)g(t,\,s_\ast)\delta(\omega-w)\delta(\omega_\ast-w)}\,ds\,ds_\ast\,d\omega\,d\omega_\ast
\end{align*}
for an arbitrary macroscopic observable $\varphi$ depending only on the headway $s$. From here, we easily check that
$$ g^\infty_{\rho,w}(s)=\delta(s-s_d(\rho,\,w)) $$
is a possible equilibrium distribution, which, consistently with the discussion set forth above, has mean $s_d(\rho,\,w)$. To prove that this is actually the only possible equilibrium distribution, at least in the quasi-invariant regime, we perform a quasi-invariant scaling inspired by that of Section~\ref{sect:local_Maxwellian.uncontrolled}. In particular, we assume~\eqref{eq:V.small_epsilon} and we observe that in order for interactions~\eqref{eq:controlled_interactions} to be quasi-invariant we also need to ensure that the additional term proportional to $s_d(\rho,\,w)-s$ gives a small contribution when the scaling parameter $\epsilon$ is small. To this end, we may further scale either $\nu=\frac{1}{\epsilon}$ or $p=\epsilon$. In both cases, letting $\varphi(s)=s^2$ we find that the trend of the energy in the quasi-invariant limit $\epsilon\to 0^+$ is ruled exactly by~\eqref{eq:E}, which, together with~\eqref{eq:h.sd}, implies $E\to s_d^2(\rho,\,w)$ for $t\to +\infty$.

In conclusion, the local Maxwellian that we consider is
$$ M_{\rho,w}(s,\,\omega)=\rho\delta(s-s_d(\rho,\,w))\otimes\delta(\omega-w). $$
Notice that in this case it is parametrised only by the hydrodynamic quantities $\rho,\,w$. As a consequence, from the transport step~\eqref{eq:transport_step} we expect a \textit{second order} macroscopic traffic model with state variables $\rho,\,w$. Indeed, proceeding like in Section~\ref{sect:macro_eq.uncontrolled} with $\phi(s,\,\omega)=1,\,\omega$ (the collisional invariants) we end up with
\begin{equation}
	\begin{cases}
 		\partial_t\rho+\partial_x\!\left(\rho V\!\left(\dfrac{1}{s_d(\rho,\,w)},\,w\right)\right)=0 \\[4mm]
 		\partial_t(\rho w)+\partial_x\!\left(\rho wV\!\left(\dfrac{1}{s_d(\rho,\,w)},\,w\right)\right)=0,
 	\end{cases}
 	\label{eq:GSOM}
\end{equation}
namely a Generic Second Order Model (GSOM) of the type introduced in~\cite{appert-rolland2009BOOK,lebacque2007PROCEEDINGS}.

A few remarks about model~\eqref{eq:GSOM} are in order. First, it is strictly hyperbolic provided $\partial_\rho s_d\neq 0$ and complies with the Aw-Rascle consistency condition if $\partial_\rho s_d\leq 0$, indeed its eigenvalues are
$$ \lambda_1=V\!\left(\frac{1}{s_d(\rho,\,w)},\,w\right)+\partial_sV\!\left(\frac{1}{s_d(\rho,\,w)},\,w\right)\partial_\rho s_d(\rho,\,w),
	\qquad \lambda_2=V\!\left(\frac{1}{s_d(\rho,\,w)},\,w\right). $$
Second, we stress again that, unlike~\eqref{eq:thirdorderwithoutcontrol} and despite the analogous derivation, it is a second order model, the ultimate reason being that the introduction of the control in the microscopic interactions destroys the local conservation of the mean headway. In particular, when locally in equilibrium the mean headway becomes a function of $\rho,\,w$, thus it no longer enters the macroscopic equations. Interestingly, the kinetic derivation of the hydrodynamic models~\eqref{eq:thirdorderwithoutcontrol},~\eqref{eq:GSOM} unveils the microscopic origin of their structural differences. Third, we observe that the penetration rate $p$ of the driver-assist vehicles does not appear explicitly in~\eqref{eq:GSOM}. The reason is again linked to the non-conservation of the local mean headway: as~\eqref{eq:h.sd} shows, $p$ affects the rate of convergence of $h$ to its local equilibrium but not the local equilibrium itself. However, it is clear that the time scale separation between local interactions and transport, which is at the basis of the hydrodynamic limit leading to~\eqref{eq:GSOM}, is more or less valid depending on the speed of convergence of the interactions to the local equilibrium. Thus,~\eqref{eq:GSOM} is implicitly valid only for a sufficiently high penetration rate $p$. In other words, it describes universal macroscopic trends of a traffic stream with a large enough percentage of driver-assist vehicles. Fourth, with the particular choice
\begin{equation}
	s_d(\rho,\,w)=\frac{1}{\rho},
	\label{eq:sd=1/rho}
\end{equation}
which satisfies $\partial_\rho s_d<0$ and reflects the usual relationship empirically assumed between the local mean headway and the traffic density, cf. e.g.,~\cite{gazis1961OR}, we obtain
\begin{equation}
	\begin{cases}
 		\partial_t\rho+\partial_x\!\left(\rho V(\rho,\,w)\right)=0 \\
 		\partial_t(\rho w)+\partial_x\!\left(\rho wV(\rho,\,w)\right)=0,
 	\end{cases}
	\label{eq:GARZ}
\end{equation}
i.e. the Generalised Aw-Rascle-Zhang (GARZ) model proposed in~\cite{fan2014NHM}. Apart from this particular case, the design of the recommended headway $s_d(\rho,\,w)$ will be the specific object of the next section.
 
\section{Hydrodynamic optimisation}
\label{sect:hydro_opt}
The recommended headway $s_d$ appears in the hydrodynamic model~\eqref{eq:GSOM} in consequence of the feedback control~\eqref{eq:u^ast} implemented in the \textit{microscopic} interaction rules~\eqref{eq:interactionscontrol} and subsequently upscaled via the Enskog-type kinetic description. The idea is now to understand the function $s_d(\rho,\,w)$ as a \textit{control} in the hydrodynamic equations and to design it so as to optimise \textit{macroscopic} traffic trends, such as the global flux or the global congestion of the vehicles. This corresponds to a \textit{multiscale} traffic control, which is explicitly implemented at the scale of single vehicles and finally produces a hydrodynamic optimisation.

\begin{remark}
In this work, we do not investigate the local mesoscopic (statistical) effects produced by a generic $s_d$. Instead, we refer the interested readers to~\cite{piccoli2019ZAMP_preprint,tosin2019MMS} for thorough analyses of this aspect.
\end{remark}

Assume that the space domain of~\eqref{eq:GSOM} is the interval $[-L,\,L]$, $L>0$, with periodic boundary conditions. This simulates a circular track, a setting often used in real experiments on traffic flow~\cite{stern2018TRC,sugiyama2008NJP}. We consider the following macroscopic functionals to be optimised:
\begin{enumerate}[label=\roman*)]
\item to maximise the global flux of vehicles we look for a control $u=u(t,\,x)$ which maximises
\begin{equation}
	J_{\rho V}(u):=\int_0^T\int_{-L}^L\left(\rho V\!\left(\frac{1}{u},\,w\right)-\mu F(u)\right)dx\,dt
	\label{eq:JrhoV}
\end{equation}
subject to~\eqref{eq:GSOM}, where, as anticipated, we understand $s_d(\rho,\,w)$ as $u$. Notice that, once determined as $u^\ast:=\operatorname{arg\,max}J_{\rho V}(u)$, the optimal control $u^\ast$ will be expressed in feedback form as a function of $\rho,\,w$, thus it will be suited to play the role of the recommended headway $s_d(\rho,\,w)$;
\item to minimise the global traffic congestion we look for a control $u=u(t,\,x)$ which minimises
\begin{equation}
	J_{\rho}(u):=\int_0^T\int_{-L}^L\left(\rho^\alpha+\mu F(u)\right)dx\,dt
	\label{eq:Jrho}
\end{equation}
subject to~\eqref{eq:GSOM} with the same relationship between $u$ and $s_d(\rho,\,w)$ set forth above. This time, however, the optimal control is determined as $u^\ast=\operatorname{arg\,min}J_{\rho}(u)$.
\end{enumerate}

In both~\eqref{eq:JrhoV} and~\eqref{eq:Jrho} $T>0$ is a finite time horizon for the optimisation, $F(u)$ is a convex penalisation function (cost of the control) and $\mu>0$ is a proportionality parameter. Furthermore, in~\eqref{eq:Jrho} the exponent $\alpha>0$ is a parameter which stresses locally high and low density regimes.

Since $u$ represents $s_d(\rho,\,w)$, the admissible controls are non-negative functions: $u(t,\,x)\geq 0$ for all $t\geq 0$ and all $x\in [-L,\,L]$. Therefore, the optimisation of the functionals $J_{\rho V}$ and $J_\rho$ should be performed under the further constraint $u\geq 0$, which however typically increases the technicality of the problem with no particular added value to the model itself. For this reason, we prefer to take into account the non-negativity of the control by choosing a penalisation function defined only for $u\geq 0$, so that on the whole both functionals~\eqref{eq:JrhoV},~\eqref{eq:Jrho} are not defined for $u<0$. A convex function $F$ complying with this requirement is
\begin{equation}
	F(u)=u\left(\log{u}-1\right)+1,
	\label{eq:F}
\end{equation}
which is also continuous on $\R_+$ up to letting $F(0):=\lim_{u\to 0^+}F(u)=1$ and such that $F'(u)=\log{u}$ for $u>0$.

\subsection{Instantaneous control}
\label{sect:inst_contr}
Consistently with the instantaneous response of the driver-assist vehicles to the actions of the human drivers, it is reasonable to understand the recommended headway as an \textit{instantaneous control strategy}. In other words, $s_d(\rho,\,w)$ should be defined in terms of the instantaneous values of $\rho,\,w$, that a driver-assist vehicle can readily detect and use, rather than on their time history over a long time horizon.

We implement this idea by considering first the functional~\eqref{eq:JrhoV}. Let $\Delta{t}>0$ be a small time interval and let us consider the following discrete-in-time version of~\eqref{eq:JrhoV} over a time horizon $[t,\,t+\Delta{t}]$:
\begin{equation}
	J_{\rho V}(u)=\Delta{t}\int_{-L}^L\left(\rho(t+\Delta{t},\,x)V\!\left(\frac{1}{u(t,\,x)},\,w(t+\Delta{t},\,x)\right)-\mu F(u(t,\,x))\right)dx
	\label{eq:JrhoV.discr}
\end{equation}
subject to the following discrete-in-time version of~\eqref{eq:GSOM}:
\begin{equation}
	\begin{cases}
 		\rho(t+\Delta{t},\,x)=\rho(t,\,x)-\Delta{t}\partial_x\!\left(\rho(t,\,x)V\!\left(\dfrac{1}{u(t,\,x)},\,w(t,\,x)\right)\right) \\[4mm]
 		w(t+\Delta{t},\,x)=w(t,\,x)-\Delta{t}V\!\left(\dfrac{1}{u(t,\,x)},\,w(t,\,x)\right)\partial_xw(t,\,x).
 	\end{cases}
 	\label{eq:GSOM_discrete}
\end{equation}
Plugging these values of $\rho(t+\Delta{t},\,x)$, $w(t+\Delta{t},\,x)$ into~\eqref{eq:JrhoV.discr} we obtain
$$ J_{\rho V}(u)=\Delta{t}\int_{-L}^L\left(\rho V\!\left(\frac{1}{u},\,w\right)-\mu F(u)\right)dx+o(\Delta{t}), $$
where we have omitted the variables $(t,\,x)$ of the quantities $\rho,\,w,\,u$ for brevity. Here, $o(\Delta{t})$ denotes higher order terms in $\Delta{t}$ that we may formally neglect under the assumption of small time horizon. To find the optimality condition associated with the maximisation of $J_{\rho V}$ we consider $u=u^\ast+\varepsilon v$, where $u^\ast$ is the (unknown) optimal control, $v$ is an arbitrary test function and $\varepsilon>0$ is a parameter. Imposing the stationarity of $J_{\rho V}$ at $u^\ast$:
$$ \left.\frac{d}{d\varepsilon}J_{\rho V}(u^\ast+\varepsilon v)\right\vert_{\varepsilon=0}=0, $$
we find the equation
$$ \int_{-L}^L\left(\rho\partial_sV\!\left(\frac{1}{u^\ast},\,w\right)-\mu F'(u^\ast)\right)v\,dx+o(1)=0, $$
which in the limit $\Delta{t}\to 0^+$ and owing to the arbitrariness of $v$ implies
\begin{equation}
	\rho\partial_sV\!\left(\frac{1}{u^\ast},\,w\right)-\mu F'(u^\ast)=0.
	\label{eq:JrhoV.optimal_u.gen}
\end{equation}
From~\eqref{eq:JrhoV.optimal_u.gen}, solving for $u^\ast$ we get the instantaneously optimal control in terms of $\rho$, $w$, which represents the recommended headway $s_d(\rho,\,w)$ for the maximisation of the flux of vehicles. For instance, if $V,\,F$ are given respectively by~\eqref{eq:V.FTL},~\eqref{eq:F} we obtain
\begin{equation}
	{(a+u^\ast)}^2\log{u^\ast}=\frac{a}{\mu}\rho w,
	\label{eq:JrhoV.optimal_u.partic}
\end{equation}
which admits a unique solution $u^\ast\geq 1$ because the left-hand side is one-to-one and onto as a function of $u$ from $\R_+$ to $\R$.

Let us repeat now these arguments for the functional~\eqref{eq:Jrho}. Its discrete-in-time version over a time horizon $[t,\,t+\Delta{t}]$ with $\Delta{t}>0$ small is
$$ J_\rho(u)=\Delta{t}\int_{-L}^L(\rho^\alpha(t+\Delta{t},\,x)+\mu F(u(t,\,x)))\,dx $$
subject to~\eqref{eq:GSOM_discrete}. Using these constraints we determine in particular
\begin{align*}
	\rho^\alpha(t+\Delta{t},\,x) &= \rho^\alpha(t,\,x)-\alpha\Delta{t}\rho^{\alpha-1}(t,\,x)\partial_x\!\left(\rho(t,\,x)V\!\left(\frac{1}{u(t,\,x)},\,w(t,\,x)\right)\right)+o(\Delta{t}) \\
	&= \rho^\alpha(t,\,x)-\alpha\Delta{t}\partial_x\!\left(\rho^\alpha(t,\,x)V\!\left(\frac{1}{u(t,\,x)},\,w(t,\,x)\right)\right) \\
	&\phantom{=} +(\alpha-1)V\!\left(\frac{1}{u(t,\,x)},\,w(t,\,x)\right)\partial_x\rho^\alpha(t,\,x)+o(\Delta{t})
\end{align*}
and we observe that $\partial_x\!\left(\rho^\alpha V\!\left(\frac{1}{u},\,w\right)\right)$ integrates to zero on $[-L,\,L]$ because of the periodic boundary conditions. Hence we obtain
$$ J_\rho(u)=\Delta{t}\int_{-L}^L\left(\rho^\alpha+(\alpha-1)\Delta{t}V\!\left(\frac{1}{u},\,w\right)\partial_x\rho^\alpha+\mu F(u)\right)dx+o(\Delta{t}^2), $$
which, imposing
$$ \left.\frac{d}{d\varepsilon}J_\rho(u^\ast+\varepsilon v)\right\vert_{\varepsilon=0}=0 $$
for an arbitrary test function $v$, produces the optimality condition
$$ \int_{-L}^L\left((\alpha-1)\Delta{t}\partial_sV\!\left(\frac{1}{u^\ast},\,w\right)\partial_x\rho^\alpha+\mu F'(u^\ast)\right)v\,dx+o(\Delta{t})=0. $$
If the penalisation coefficient $\mu$ is independent of $\Delta{t}$ then in the limit $\Delta{t}\to 0^+$ we get $F'(u^\ast)=0$, namely an equation for the optimal control independent of $\rho$, $w$. If instead we scale the penalisation coefficient as $\mu=\kappa\Delta{t}$ with $\kappa>0$, meaning that the cost of the control is proportional to the length of the time horizon of the optimisation, then for $\Delta{t}\to 0^+$ we have
\begin{equation}
	(\alpha-1)\partial_sV\!\left(\frac{1}{u^\ast},\,w\right)\partial_x\rho^\alpha+\kappa F'(u^\ast)=0,
	\label{eq:Jrho.optimal_u.gen}
\end{equation}
whence we get in general a richer instantaneous optimal control, viz. recommended headway $s_d$, depending on $\rho$, $w$. Notice however that for $\alpha=1$, i.e. if the goal is to minimise $\rho$ itself,~\eqref{eq:Jrho.optimal_u.gen} reduces in turn to $F'(u^\ast)=0$. On the other hand, if $\alpha>0$ is generic and $V,\,F$ are given by~\eqref{eq:V.FTL},~\eqref{eq:F} then~\eqref{eq:Jrho.optimal_u.gen} yields
\begin{equation}
	{(a+u^\ast)}^2\log{u^\ast}=(1-\alpha)\frac{a}{\kappa}w\partial_x\rho^\alpha,
	\label{eq:Jrho.optimal_u.partic}
\end{equation}
which admits a unique solution $u^\ast\geq 0$. In particular, for $\alpha=1$ this solution is $u^\ast=1$, viz. a constant unitary recommended headway.

\subsection{Application to the Aw-Rascle-Zhang model}
The Aw-Rascle-Zhang (ARZ) model is a very popular traffic model of the form~\eqref{eq:GARZ} with
$$ V(\rho,\,w)=w-p(\rho), $$
$p:\R_+\to\R_+$ being a monotonically increasing function called the \textit{traffic pressure}. This model was proposed by Aw and Rascle~\cite{aw2000SIAP}, and independently by Zhang~\cite{zhang2002TRB}, to overcome some drawbacks of second order macroscopic traffic models pointed out by Daganzo~\cite{daganzo1995TR}. The traffic pressure is usually taken of the form
\begin{equation}
	p(\rho)=\rho^\delta, \qquad \delta>0.
	\label{eq:p}
\end{equation}

Recalling that, in the present context,~\eqref{eq:GARZ} is obtained from~\eqref{eq:GSOM} with the choice~\eqref{eq:sd=1/rho}, we can recast the ARZ model in the controlled setting~\eqref{eq:GSOM} by letting
\begin{equation}
	V\!\left(\frac{1}{s},\,\omega\right)=\omega-p\!\left(\frac{1}{s}\right),
	\label{eq:V.AR}
\end{equation}
then we can exploit the results of Section~\ref{sect:inst_contr} to deduce instantaneous optimal controls for flux maximisation and congestion minimisation.

Specifically, condition~\eqref{eq:JrhoV.optimal_u.gen} for the maximisation of the flux becomes
$$ \frac{\rho}{{(u^\ast)}^2}p'\!\left(\frac{1}{u^\ast}\right)-\mu F'(u^\ast)=0, $$
which for $F,\,p$ like in~\eqref{eq:F},~\eqref{eq:p} produces
\begin{equation}
	{(u^\ast)}^{1+\delta}\log{u^\ast}=\frac{\delta}{\mu}\rho.
	\label{eq:JrhoV.optimal_u.AR}
\end{equation}
This equation admits a unique solution $u^\ast\geq 1$ because the left-hand side is one-to-one and onto as a function of $u$ from $[1,\,+\infty)$ to $\R_+$. Notice that the resulting recommended headway $s_d(\rho,\,w)=u^\ast$ is actually independent of $w$.

On the other hand, condition~\eqref{eq:Jrho.optimal_u.gen} for the minimisation of the traffic congestion becomes
$$ \frac{\alpha-1}{{(u^\ast)}^2}p'\!\left(\frac{1}{u^\ast}\right)\partial_x\rho^\alpha+\kappa F'(u^\ast)=0, $$
which with $F,\,p$ like in~\eqref{eq:F},~\eqref{eq:p} yields
$$ {(u^\ast)}^{1+\delta}\log{u^\ast}=(1-\alpha)\frac{\delta}{\kappa}\partial_x\rho^\alpha. $$
This equation is ill posed if $(1-\alpha)\partial_x\rho^\alpha\leq 0$. Indeed, the mapping $u\mapsto u^{1+\delta}\log{u}$ is decreasing for $0<u<e^{-\frac{1}{1+\delta}}$, increasing for $u>e^{-\frac{1}{1+\delta}}$ and reaches the absolute minimum $-\frac{1}{(1+\delta)e}$ at $u=e^{-\frac{1}{1+\delta}}$. Consequently, if $-\frac{1}{(1+\delta)e}\leq (1-\alpha)\partial_x\rho^\alpha\leq 0$ there are two solutions whereas if $(1-\alpha)\partial_x\rho^\alpha<-\frac{1}{(1+\delta)e}$ there is no solution.

\begin{remark}
The speed function~\eqref{eq:V.AR}, together with the choice~\eqref{eq:p} of the traffic pressure, complies neither with Assumption~\ref{ass:V}\ref{item:V.bound.lin} nor with~\eqref{eq:V.small_epsilon}. Therefore, the inclusion of the ARZ model among the particular cases obtainable from~\eqref{eq:GSOM} is only formal, being not strictly supported by the derivation performed in Sections~\ref{sect:h.o._macro_derivation},~\ref{sect:GSOM}. We point out that a genuine Enskog-type kinetic derivation of the ARZ model with uncontrolled speed-based vehicle interactions may instead be found in the recent paper~\cite{dimarco2020JSP}.
\end{remark}

\section{Numerical tests}
\label{sect:numerics}
We exemplify now the results of Section~\ref{sect:hydro_opt} through selected numerical tests. In more detail, we solve numerically the hydrodynamic model~\eqref{eq:GSOM} with $s_d$ chosen out of the instantaneous optimisation of either functional~\eqref{eq:JrhoV},~\eqref{eq:Jrho} and we compare the results with those obtained by fixing \textit{a priori} $s_d$ like in~\eqref{eq:sd=1/rho}, which produces the GARZ model~\eqref{eq:GARZ}.

We consider both the speed function~\eqref{eq:V.FTL}, motivated by FTL microscopic dynamics, and the speed function~\eqref{eq:V.AR}, directly suggested by the ARZ macroscopic model.

In all cases, we solve the hydrodynamic model by means of an upwind scheme coupled with a non-linear algebraic solver of~\eqref{eq:JrhoV.optimal_u.gen},~\eqref{eq:Jrho.optimal_u.gen} at each grid point $(x,\,t)$. Consistently with the theory developed in Section~\ref{sect:hydro_opt}, we take as spatial domain the interval $[-1,\,1]$ with periodic boundary conditions, which simulates a circular track. As initial conditions $\rho_0(x):=\rho(0,\,x)$, $w_0(x):=w(0,\,x)$, we prescribe
$$ \rho_0(x)=
	\begin{cases}
		0.8 & \text{if } x\leq 0 \\
		0 & \text{if } x>0,
		%10^{-3} & \text{if } x>0,
	\end{cases}
	\qquad
	w_0(x)=
	\begin{cases}
		0.55 & \text{if } x\leq 0 \\
		0.5 & \text{if } x>0,
	\end{cases}
$$
which mimic a platoon of vehicles filling initially one half of the circular track.% The small though non-zero value of $\rho_0$ for $x>0$ is meant to describe a substantial absence of vehicles in that portion of the domain, avoiding at the same time the drawbacks caused in second order macroscopic models by vacuum states.

\begin{figure}[!t]
\centering
\includegraphics[width=\textwidth]{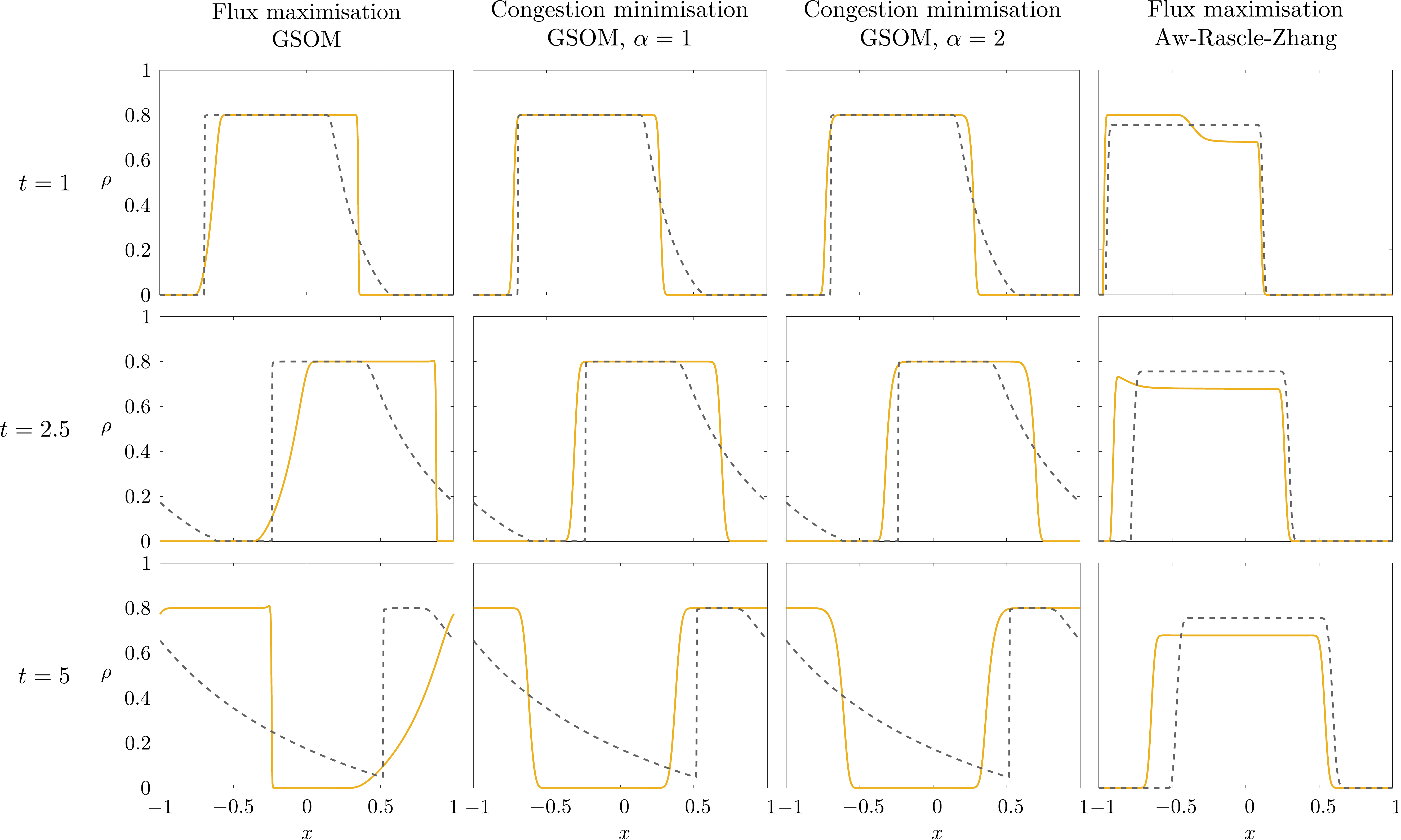}
\caption{Density profiles at three successive computational times obtained from the numerical solution of the GSOM~\eqref{eq:GSOM} with the speed function~\eqref{eq:V.FTL} (first three columns from the left) and the speed function~\eqref{eq:V.AR} (fourth column from the left). Solid lines: optimal choice of $s_d$ for the optimisations indicated on the top of the columns. Dashed lines: ``standard'' choice $s_d=\frac{1}{\rho}$, cf.~\eqref{eq:sd=1/rho}.}
\label{fig:densities}
\end{figure}

The first three columns from the left of Figure~\ref{fig:densities} show the density profiles (solid lines) at the three successive computational times $t=1,\,2.5,\,5$ obtained with the GSOM~\eqref{eq:GSOM} with $V$ given by~\eqref{eq:V.FTL} in the cases of flux maximisation and congestion minimisation. The flux maximisation (first column) is ruled by the optimality condition~\eqref{eq:JrhoV.optimal_u.partic} with $\mu=0.1$ and $a=1$ whereas the congestion minimisation (second and third columns) is ruled by~\eqref{eq:Jrho.optimal_u.partic} with $\kappa=0.1$, $a=1$ and $\alpha=1,\,2$. The dashed line is instead the density profile obtained from~\eqref{eq:GSOM} with $s_d$ given by~\eqref{eq:sd=1/rho}, i.e. with no specific optimisation. It is clear that the optimal $s_d$'s operate so as to keep the platoon of vehicles compact. In particular, they avoid the formation of a rarefaction wave responsible for the spreading of the density across the whole domain.
\begin{figure}[!t]
\centering
\subfigure[GSOM]{\includegraphics[width=\textwidth]{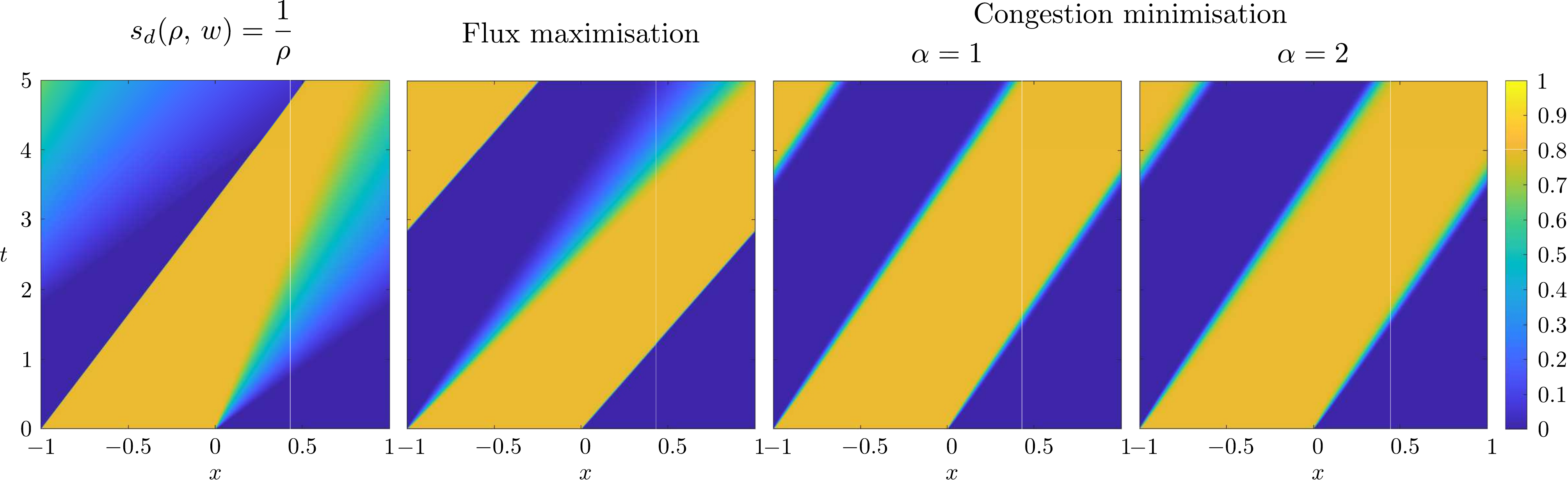}}
\subfigure[Aw-Rascle-Zhang]{\includegraphics[width=0.55\textwidth]{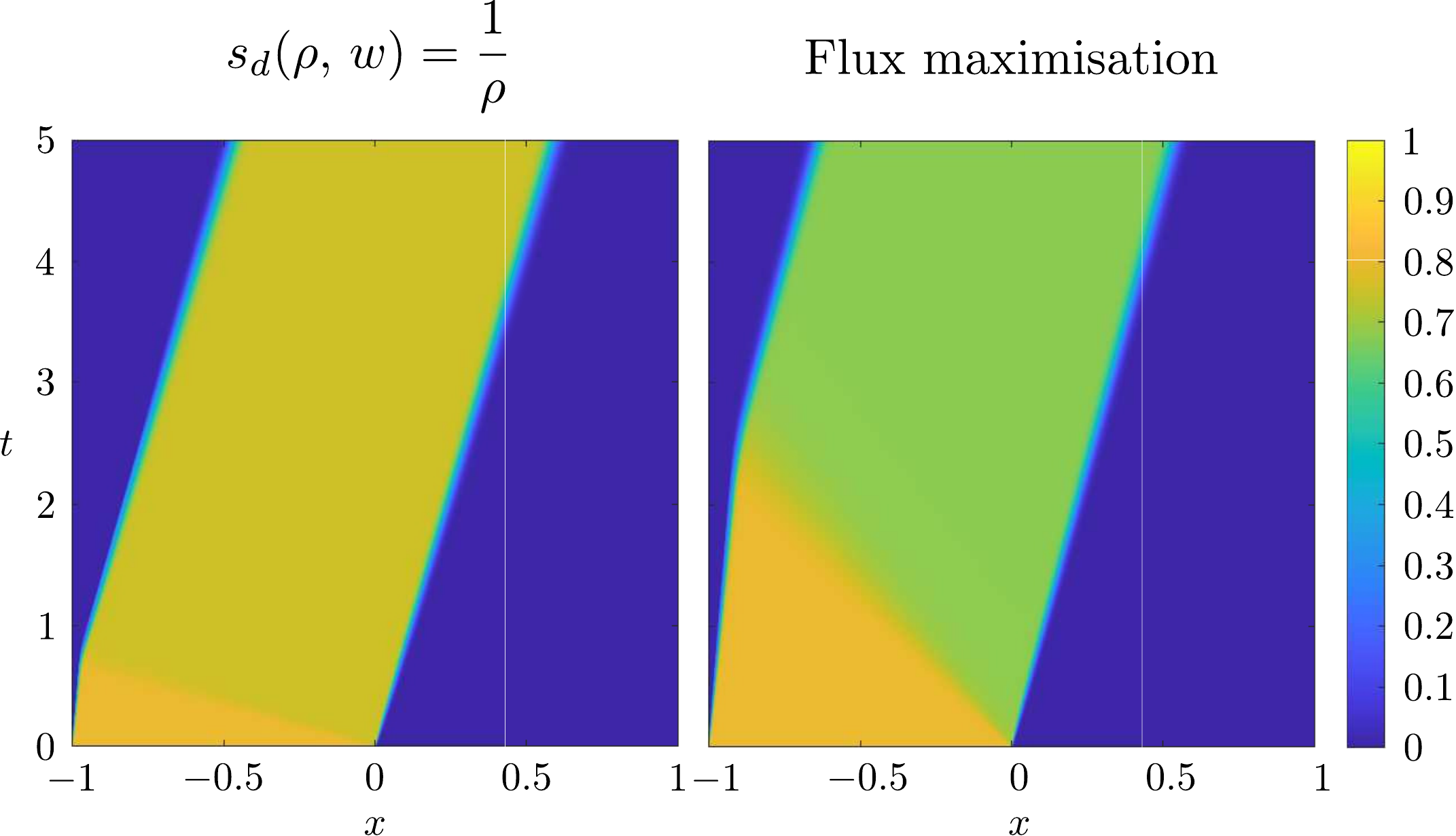}}
\caption{Wave diagrams in the $xt$-plane corresponding to: (a) the first three columns from the left of Figure~\ref{fig:densities}; (b) the fourth column from the left of Figure~\ref{fig:densities}.}
\label{fig:tx}
\end{figure}
This effect is further emphasised by the wave diagrams in the $xt$-plane shown in Figure~\ref{fig:tx}(a).
\begin{figure}[!t]
\centering
\subfigure[GSOM]{\includegraphics[width=\textwidth]{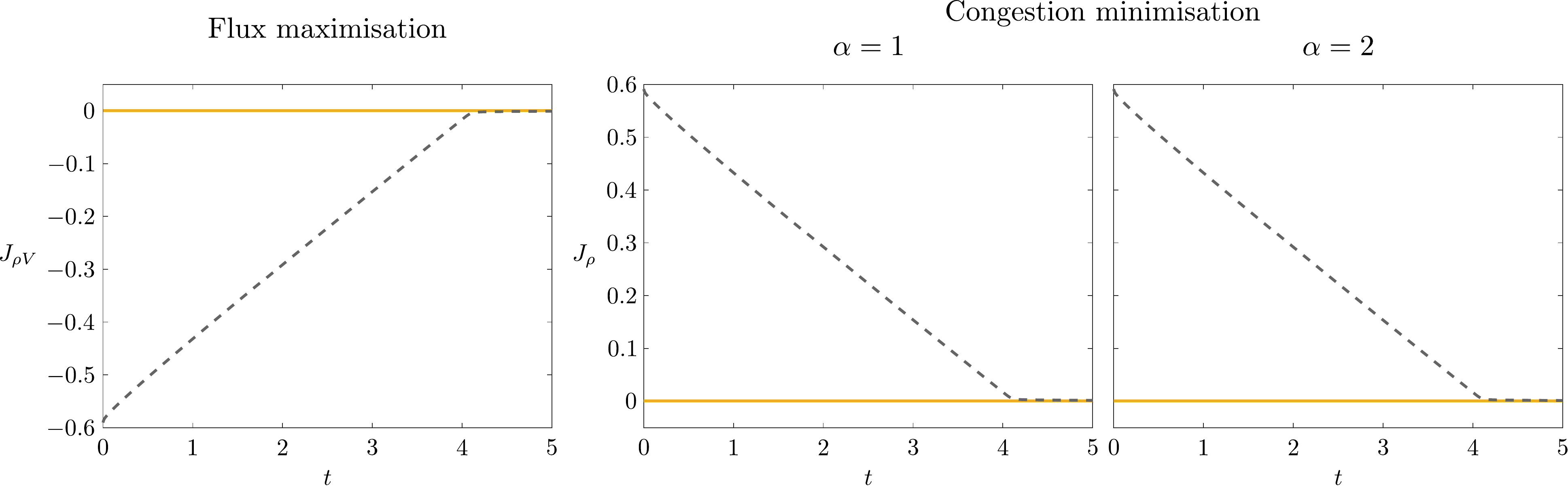}}
\subfigure[Aw-Rascle-Zhang]{\includegraphics[width=0.36\textwidth]{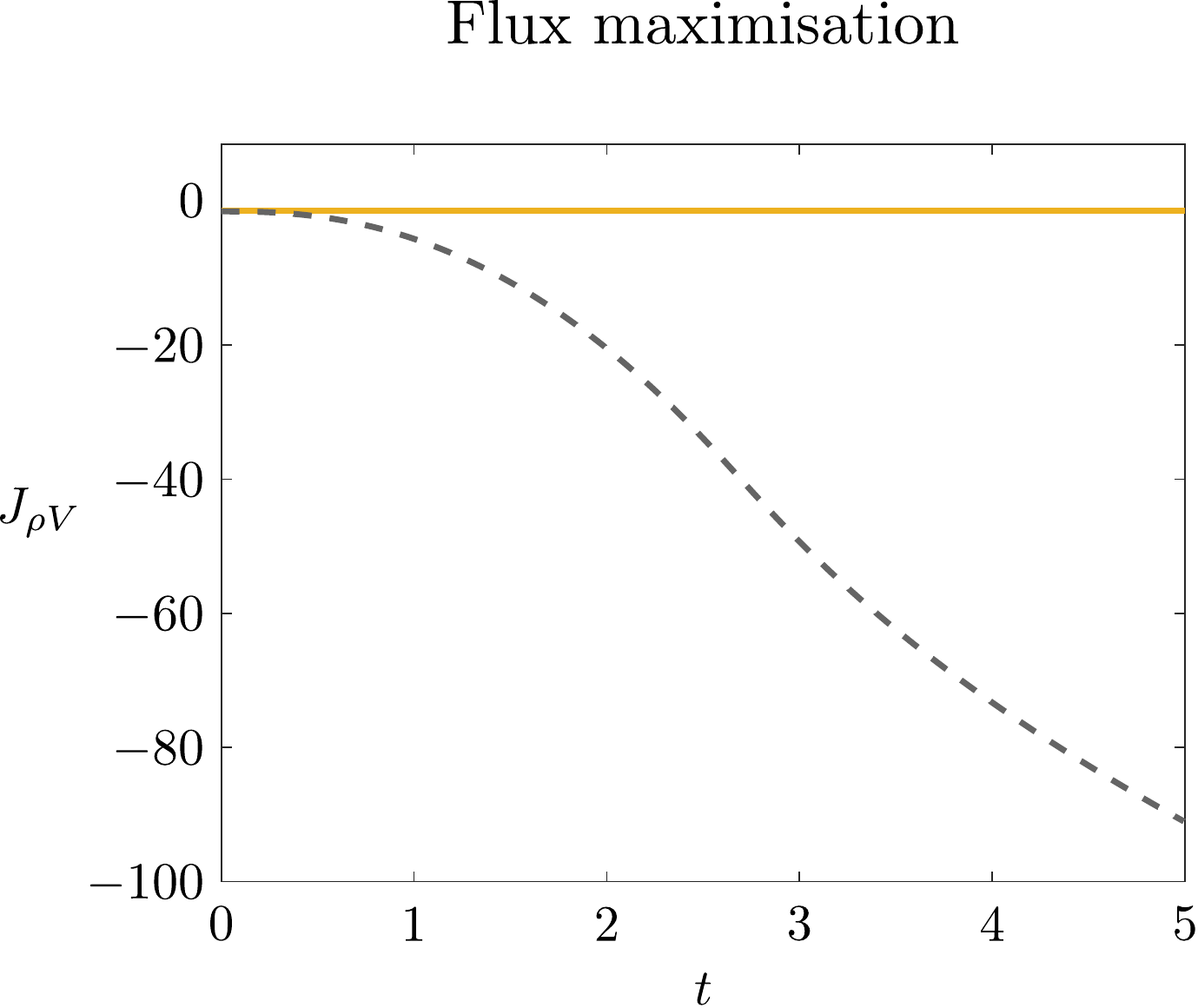}}
\caption{Instantaneous values of the functionals $J_{\rho V}$, $J_\rho$ corresponding to: (a) the first three columns from the left of Figure~\ref{fig:densities} and to Figure~\ref{fig:tx}(a); (b) the fourth column from the left of Figure~\ref{fig:densities} and to Figure~\ref{fig:tx}(b). Solid lines: optimal choice of $s_d$ for the optimisations indicated on the top of the pictures. Dashed lines: ``standard'' choice $s_d=\frac{1}{\rho}$, cf.~\eqref{eq:sd=1/rho}.}
\label{fig:J}
\end{figure}
Finally, Figure~\ref{fig:J}(a) shows the instantaneous values of the functionals $J_{\rho V}$, $J_\rho$ with the optimal $s_d$'s (solid line) and with $s_d$ given by~\eqref{eq:sd=1/rho} (dashed line). It is interesting to observe that, starting approximately from the computational time $t=4$, the functionals take the same values both in the optimised and in the non-optimised cases. This is probably a consequence of the periodic boundary conditions, which, in the long run, tend to make the \textit{integral} values of the flux and the density uniform despite persisting differences in the corresponding \textit{pointwise} profiles.

The fourth column from the left of Figure~\ref{fig:densities} compares the density profiles with (solid line) and without (dashed line) flux maximisation obtained from the ARZ model, i.e. the GSOM~\eqref{eq:GSOM} with $V$ given now by~\eqref{eq:V.AR}. In this case, the flux maximisation is ruled by~\eqref{eq:JrhoV.optimal_u.AR} with $\mu=0.1$ and $\delta=3$ while the non-optimised case is again obtained taking $s_d$ like in~\eqref{eq:sd=1/rho}. We observe that the flux maximisation is achieved through a redistribution of the vehicles in the platoon. Initially they are slowed down, whereby their density diminishes and the rear part of the platoon elongates. Subsequently, the platoon remains compact and recovers essentially the same speed as in the non-optimised case, cf. the wave diagrams in Figure~\ref{fig:tx}(b). From Figure~\ref{fig:J}(b) we observe that, unlike the previous cases, the instantaneous values of the non-optimised functional $J_{\rho V}$ (dashed line) depart more and more consistently from those of the optimised one (solid line), probably as a consequence of a much higher implementation cost (viz. penalisation) of the non-optimal $s_d$.

\section{Conclusions}
\label{sect:conclusions}
In this paper, we have derived generic high order macroscopic traffic models from a feedback-controlled particle description via an Enskog-type kinetic approach.

At the microscopic scale, we have considered a class of generic Follow-the-Leader (FTL) models which include a Lagrangian marker, i.e. a label attached to each vehicle representing a constant-in-time driving characteristic, such as e.g., the maximum speed. We have shown that the corresponding natural macroscopic description is provided by a \textit{third order} hyperbolic system of conservation/balance laws for the density of vehicles, their mean Lagrangian marker and the mean headway among them. These are the hydrodynamic parameters conserved by the FTL interactions, or in classical kinetic terms the ``collisional'' invariants.

Next, we have included a feedback control in the FTL interaction rules, which mimics the action of a driver-assistance system trying to maintain a recommended distance $s_d$ from the leading vehicle. We have modelled $s_d$ as a parameter depending on the local traffic congestion and the local mean Lagrangian marker. Moreover, we have taken into account that all vehicles may not be equipped with such a controller. For this, we have assumed that a randomly selected vehicle is controlled with a certain probability $p$ understood as the penetration rate of the driver-assist technology in the traffic stream. In the regime of sufficiently high $p$, we have upscaled the controlled FTL model to a macroscopic model by taking the hydrodynamic limit of the corresponding Enskog-type kinetic description.

We have shown that the resulting hydrodynamic model describes universal traffic trends for large enough penetration rates. Indeed, $p$ does not parametrise the macroscopic equations but affects the convergence rate of the microscopic interactions to their local equilibrium. Remarkably, this hydrodynamic model turns out to be a \textit{second order} one belonging to the GSOM class. The order reduction with respect to the uncontrolled case has its origin in the fact that the introduction of the driver-assist control destroys the local conservation of the mean headway among the vehicles. Furthermore, this model is parametrised by the recommended distance $s_d$, which we have proposed to understand as a further control to be fixed in such a way to optimise macroscopic traffic dynamics. Using the technique of the instantaneous control, which is particularly meaningful for driver-assist vehicles, we have proved that there exist instantaneously optimal choices of $s_d$ (i.e. optimal $s_d$'s based on the instantaneous values of the hydrodynamic variables describing the traffic stream) which e.g., maximise the flow of vehicles or minimise the traffic congestion. Apart from these two examples, the technique that we have proposed is quite general and may also be applied to other macroscopic functionals to be optimised.

Summarising, in this paper we have ultimately performed a \textit{multiscale control} and \textit{optimisation} of traffic. Indeed, starting from a microscopic control, which optimises the interaction of a single vehicle with its leading vehicle, we have shown that it is possible to design explicitly the control parameters so as to optimise global traffic trends. This also suggests that vehicle-wise automatic decision algorithms may successfully turn driver-assist vehicles into \textit{bottom-up traffic controllers}, provided their penetration rate in the traffic stream is sufficiently high. On the other hand, we believe that the conceptual scheme we have proposed in this paper may be fruitfully applied also to the multiscale control of several other multi-agent systems, such as e.g., human crowds or social systems, in which desired collective trends cannot be simply obtained by top-down impositions but need rather to emerge spontaneously from suitably controlled individual interactions.

\section*{Acknowledgements}
This research was partially supported  by the Italian Ministry for Education, University and Research (MIUR) through the ``Dipartimenti di Eccellenza'' Programme (2018-2022), Department of Mathematical Sciences ``G. L. Lagrange'', Politecnico di Torino (CUP: E11G18000350001) and through the PRIN 2017 project (No. 2017KKJP4X) ``Innovative numerical methods for evolutionary partial differential equations and applications''.

F.A.C. acknowledges support from ``Compagnia di San Paolo'' (Torino, Italy)

F.A.C. and A.T. are members of GNFM (Gruppo Nazionale per la Fisica Matematica) of INdAM (Istituto Nazionale di Alta Matematica), Italy.

The research of B.P. is based upon work supported by the U.S. Department of Energy's Office of Energy Efficiency and Renewable Energy (EERE) under the Vehicle Technologies Office award number CID DE-EE0008872. The views expressed herein do not necessarily represent the views of the U.S. Department of Energy or the United States Government.

\bibliographystyle{plain}
\bibliography{CfPbTa-kinetic_GSOM_control}

\begin{thebibliography}{10}

\bibitem{albi2017AMO}
G.~Albi, Y.-P. Choi, M.~Fornasier, and D.~Kalise.
\newblock Mean field control hierarchy.
\newblock {\em Appl. Math. Optim.}, 76(1):93--135, 2017.

\bibitem{appert-rolland2009BOOK}
C.~Appert-Rolland, F.~Chevoir, P.~Gondret, S.~Lassarre, J.-P. Lebacque, and
  M.~Schreckenberg, editors.
\newblock {\em Traffic and Granular Flow ’07}. Springer, 2009.

\bibitem{aw2000SIAP}
A.~Aw and M.~Rascle.
\newblock Resurrection of ``second order'' models of traffic flow.
\newblock {\em SIAM J. Appl. Math.}, 60(3):916--938, 2000.

\bibitem{carrillo2014CISM}
J.~A. Carrillo, Y.-P. Choi, and M.~Hauray.
\newblock The derivation of swarming models: mean-field limit and {W}asserstein
  distances.
\newblock In A.~Muntean and F.~Toschi, editors, {\em Collective Dynamics from
  Bacteria to Crowds}, volume 553 of {\em {CISM} {I}nternational {C}entre for
  {M}echanical {S}ciences}, pages 1--46. Springer, Vienna, 2014.

\bibitem{carrillo2010SIMA}
J.~A. Carrillo, M.~Fornasier, J.~Rosado, and G.~Toscani.
\newblock Asymptotic flocking dynamics for the kinetic {C}ucker-{S}male model.
\newblock {\em SIAM J. Math. Anal.}, 42(1):218--236, 2010.

\bibitem{carrillo2010MSSET}
J.~A. Carrillo, M.~Fornasier, G.~Toscani, and F.~Vecil.
\newblock Particle, kinetic, and hydrodynamic models of swarming.
\newblock In G.~Naldi, L.~Pareschi, and G.~Toscani, editors, {\em Mathematical
  Modeling of Collective Behavior in Socio-Economic and Life Sciences},
  Modeling and Simulation in Science, Engineering and Technology, pages
  297--336. Birkh\"{a}user, Boston, 2010.

\bibitem{chiarello2020SIAM}
F.~A. Chiarello, J.~Friedrich, P.~Goatin, and S.~G\"{o}ttlich.
\newblock Micro-{M}acro limit of a non-local generalized {A}w-{R}ascle type
  model.
\newblock {\em SIAM J. Appl. Math.}, 2020.
\newblock In press.

\bibitem{colombo2014RSMUP}
R.~M. Colombo and E.~Rossi.
\newblock On the micro-macro limit in traffic flow.
\newblock {\em Rend. Semin. Mat. Univ. Padova}, 131:217--235, 2014.

\bibitem{cordier2005JSP}
S.~Cordier, L.~Pareschi, and G.~Toscani.
\newblock On a kinetic model for a simple market economy.
\newblock {\em J. Stat. Phys.}, 120(1):253--277, 2005.

\bibitem{cristiani2011MMS}
E.~Cristiani, B.~Piccoli, and A.~Tosin.
\newblock Multiscale modeling of granular flows with application to crowd
  dynamics.
\newblock {\em Multiscale Model. Simul.}, 9(1):155--182, 2011.

\bibitem{cristiani2016NHM}
E.~Cristiani and S.~Sahu.
\newblock On the micro-to-macro limit for first-order traffic flow models on
  networks.
\newblock {\em Netw. Heterog. Media}, 11(3):395--413, 2016.

\bibitem{daganzo1995TR}
C.~F. Daganzo.
\newblock Requiem for second-order fluid approximation of traffic flow.
\newblock {\em Transportation Res.}, 29(4):277--286, 1995.

\bibitem{delis2015CMA}
A.~I. Delis, I.~K. Nikolos, and M.~Papageorgiou.
\newblock Macroscopic traffic flow modeling with adaptive cruise control:
  {D}evelopment and numerical solution.
\newblock {\em Comput. Math. Appl.}, 70(8):1921--1947, 2015.

\bibitem{delis2018TRR}
A.~I. Delis, I.~K. Nikolos, and M.~Papageorgiou.
\newblock A macroscopic multi-lane traffic flow model for {ACC}/{CACC} traffic
  dynamics.
\newblock {\em Transp. Res. Record}, 2018.

\bibitem{dellemonache2014JDE}
M.~L. Delle~Monache and P.~Goatin.
\newblock Scalar conservation laws with moving constraints arising in traffic
  flow modeling: {A}n existence result.
\newblock {\em J. Differential Equations}, 257(11):4015--4029, 2014.

\bibitem{difrancesco2017MBE}
M.~Di~Francesco, S.~Fagioli, and M.~Rosini.
\newblock Many particle approximation of the {A}w-{R}ascle-{Z}hang second order
  model for vehicular traffic.
\newblock {\em Math. Biosci. Eng.}, 14(1):127--141, 2017.

\bibitem{difrancesco2015ARMA}
M.~Di~Francesco and M.~D. Rosini.
\newblock Rigorous derivation of nonlinear scalar conservation laws from
  {F}ollow-the-{L}eader type models via many particle limit.
\newblock {\em Arch. Ration. Mech. Anal.}, 217(3):831--871, 2015.

\bibitem{dimarco2020JSP}
G.~Dimarco and A.~Tosin.
\newblock The {A}w-{R}ascle traffic model: {E}nskog-type kinetic derivation and
  generalisations.
\newblock {\em J. Stat. Phys.}, 178(1):178--210, 2020.

\bibitem{fan2014NHM}
S.~Fan, M.~Herty, and B.~Seibold.
\newblock Comparative model accuracy of a data-fitted generalized
  {A}w-{R}ascle-{Z}hang model.
\newblock {\em Netw. Heterog. Media}, 9(2):239--268, 2014.

\bibitem{fornasier2014PTRSA}
M.~Fornasier, B.~Piccoli, and F.~Rossi.
\newblock Mean-field sparse optimal control.
\newblock {\em Philos. Trans. R. Soc. A-Math. Phys. Eng. Sci.},
  372(2028):20130400/1--21, 2014.

\bibitem{garavello2020JDE}
M.~Garavello, P.~Goatin, T.~Liard, and B.~Piccoli.
\newblock A multiscale model for traffic regulation via autonomous vehicles.
\newblock {\em J. Differential Equations}, 2020.
\newblock To appear.

\bibitem{gazis1961OR}
D.~C. Gazis, R.~Herman, and R.~W. Rothery.
\newblock Nonlinear follow-the-leader models of traffic flow.
\newblock {\em Oper. Res.}, 9:545--567, 1961.

\bibitem{goatin2017CMS}
P.~Goatin and F.~Rossi.
\newblock A traffic flow model with non-smooth metric interaction:
  well-posedness and micro-macro limit.
\newblock {\em Commun. Math. Sci.}, 15(1):261--287, 2017.

\bibitem{helbing1995PRE}
D.~Helbing.
\newblock Improved fluid-dynamic model for vehicular traffic.
\newblock {\em Phys. Rev. E}, 51(4):3164--3169, 1995.

\bibitem{herty2007NHM}
M.~Herty, L.~Pareschi, and M.~Sea\"{i}d.
\newblock Enskog-like discrete velocity models for vehicular traffic flow.
\newblock {\em Netw. Heterog. Media}, 2(3):481--496, 2007.

\bibitem{klar1997JSP}
A.~Klar and R.~Wegener.
\newblock Enskog-like kinetic models for vehicular traffic.
\newblock {\em J. Stat. Phys.}, 87(1-2):91--114, 1997.

\bibitem{lattanzio2011SIMA}
C.~Lattanzio, A.~Maurizi, and B.~Piccoli.
\newblock Moving bottlenecks in car traffic flow: a {PDE}-{ODE} coupled model.
\newblock {\em SIAM J. Math. Anal.}, 43(1):50--67, 2011.

\bibitem{lebacque2007PROCEEDINGS}
J.-P. Lebacque, S.~Mammar, and H.~Haj~Salem.
\newblock Generic second order traffic flow modelling.
\newblock In R.~E. Allsop, B.~Heydecker, and M.~G.~H. Bell, editors, {\em
  Transportation and Traffic Theory 2007}, pages 755--776, 2007.

\bibitem{liard2020PREPRINT}
T.~Liard, R.~Stern, and M.~L. Delle~Monache.
\newblock A {PDE}-{ODE} model for traffic control with autonomous vehicles.
\newblock Submitted.

\bibitem{lighthill1955PRSL}
M.~J. Lighthill and G.~B. Whitham.
\newblock On kinematic waves. {II}. {A} theory of traffic flow on long crowded
  roads.
\newblock {\em Proc. Roy. Soc. London. Ser. A.}, 229:317--345, 1955.

\bibitem{ntousakis2015TRP}
I.~A. Ntousakis, I.~K. Nikolos, and M.~Papageorgiou.
\newblock On microscopic modelling of adaptive cruise control systems.
\newblock {\em Transp. Res. Procedia}, 6:111--127, 2015.

\bibitem{piccoli2019ZAMP_preprint}
B.~Piccoli, A.~Tosin, and M.~Zanella.
\newblock Model-based assessment of the impact of driver-assist vehicles using
  kinetic theory.
\newblock Submitted.

\bibitem{ramadan2020SEMAI-SIMAI}
R.~A. Ramadan, R.~R. Rosales, and B.~Seibold.
\newblock Structural properties of the stability of jamitons.
\newblock In G.~Puppo and A.~Tosin, editors, {\em Mathematical Descriptions of
  Traffic Flow: Micro, Macro and Kinetic Models}, ICIAM2019 SEMAI SIMAI
  Springer Series. Springer International Publishing, 2020.
\newblock To appear.

\bibitem{richards1956OR}
P.~I. Richards.
\newblock Shock waves on the highway.
\newblock {\em Operations Res.}, 4:42--51, 1956.

\bibitem{stern2018TRC}
R.~E. Stern, S.~Cui, M.~L. Delle~Monache, R.~Bhadani, M.~Bunting, M.~Churchill,
  N.~Hamilton, R.~Haulcy, H.~Pohlmann, F.~Wu, B.~Piccoli, B.~Seibold,
  J.~Sprinkle, and D.~B. Work.
\newblock Dissipation of stop-and-go waves via control of autonomous vehicles:
  {F}ield experiments.
\newblock {\em Transportation Res. Part C}, 89:205--221, 2018.

\bibitem{sugiyama2008NJP}
Y.~Sugiyama, M.~Fukui, M.~Kikuchi, K.~Hasebe, A.~Nakayama, K.~Nishinari,
  S.~Tadaki, and S.~Yukawa.
\newblock Traffic jams without bottlenecks -- experimental evidence for the
  physical mechanism of the formation of a jam.
\newblock {\em New J. Phys.}, 10:033001/1--7, 2008.

\bibitem{tosin2019MMS}
A.~Tosin and M.~Zanella.
\newblock Kinetic-controlled hydrodynamics for traffic models with
  driver-assist vehicles.
\newblock {\em Multiscale Model. Simul.}, 17(2):716--749, 2019.

\bibitem{tosin2020SEMA-SIMAI}
A.~Tosin and M.~Zanella.
\newblock {B}oltzmann-type description with cutoff of {F}ollow-the-{L}eader
  traffic models.
\newblock In G.~Albi, S.~Merino-Aceituno, A.~Nota, and M.~Zanella, editors,
  {\em Trails in Kinetic Theory: Foundational Aspects and Numerical Methods},
  SEMAI SIMAI Springer Series. Springer International Publishing, 2020.
\newblock To appear.

\bibitem{tosin2020MCRF}
A.~Tosin and M.~Zanella.
\newblock Uncertainty damping in kinetic traffic models by driver-assist
  controls.
\newblock {\em Math. Control Relat. Fields}, 2020.
\newblock To appear.

\bibitem{villani1998PhD}
C.~Villani.
\newblock {\em Contribution \`{a} l'\'{e}tude math\'{e}matique des
  \'{e}quations de Boltzmann et de Landau en th\'{e}orie cin\'{e}tique des gaz
  et des plasmas}.
\newblock Ph{D} thesis, Paris 9, 1998.

\bibitem{villani1998ARMA}
C.~Villani.
\newblock On a new class of weak solutions to the spatially homogeneous
  {B}oltzmann and {L}andau equations.
\newblock {\em Arch. Ration. Mech. Anal.}, 143(3):273--307, 1998.

\bibitem{zhang2002TRB}
H.~M. Zhang.
\newblock A non-equilibrium traffic model devoid of gas-like behavior.
\newblock {\em Transportation Res. Part B}, 36(3):275--290, 2002.

\end{thebibliography}
\end{document}